\documentclass[preprint,aps,prl,10pt,amsmath,amssymb,nofootinbib,showpacs,superscriptaddress,floatfix]{revtex4-1}

\newcommand{\rs}{\rm\scriptscriptstyle}

\DeclareFontFamily{U}{rcjhbltx}{}
\DeclareFontShape{U}{rcjhbltx}{m}{n}{<->rcjhbltx}{}
\DeclareSymbolFont{hebrewletters}{U}{rcjhbltx}{m}{n}

\usepackage{graphicx}
\usepackage{color}
\usepackage[usenames,dvipsnames]{xcolor}
\usepackage[colorlinks=true,linkcolor=Red,citecolor=Green,linktoc=page]{hyperref}
\usepackage{multirow}
\usepackage{float}
\usepackage{flushend}
\usepackage{balance}
\usepackage[varg]{txfonts}
\usepackage{ulem}
\usepackage{fancyhdr}

\DeclareMathSymbol{\lamed}{\mathord}{hebrewletters}{108}
\begin{document}
\title{Superinsulation, electric confinement in condensed matter \footnote{Invited review for Int. J. Mod. Phys. B}}

\author{M.\,C.\,Diamantini}

\affiliation{NiPS Laboratory, INFN and Dipartimento di Fisica e Geologia, University of Perugia, via A. Pascoli, I-06100 Perugia, Italy}

\author{C.\,A.\,Trugenberger}

\affiliation{SwissScientific Technologies SA, rue du Rhone 59, CH-1204 Geneva, Switzerland}
\affiliation{Division of Science, New York University Abu Dhabi, Abu Dhabi, United Arab Emirates}

%\author{V. M. Vinokur}
%\affiliation{Terra Quantum AG, St. Gallerstrasse 16A, 9400 Rorschach, Switzerland}
\begin{center}
    \itshape In memoriam Valerii M. Vinokur
\end{center}

%\date{\today}
\begin{abstract}
Superinsulation is the confining phase of condensed matter systems. It is a genuine quantum state of matter, dual to superconductivity, just as confinement in particle physics is dual to the Higgs phase. In contrast to hadron physics, the gauge field that becomes confining is not colour SU(3) but the conventional U(1) electromagnetic field. Superinsulation arises in superconductors with an emergent granular condensate when core-less XY vortices between condensate grains become tensionless, unobservable Dirac strings and magnetic monopoles are liberated and condense. This magnetic condensate confines electric fields into flux tubes, strings that generate a linear potential between Cooper pairs. These dual Abrikosov flux tubes give rise to infinite electrical resistance, persisting up to a finite deconfinement temperature. We shall review the theoretical framework of superinsulation together with its experimental validation, including the direct observation of the linear confining potential and we shall stress how phase-only approaches and disorder-based ideas for the phases around the superconductor-to-insulator transition are hard to reconcile with recent experimental observations. 
 
\end{abstract}
\maketitle

%\section{Introduction}
%\vspace{-0.3cm}
%\section{Introduction}

\section{Introduction}

Typical insulators are materials with a completely filled band and, consequently, a gap for charged excitations. When there are available arbitrarily low-lying charged excitations, the material becomes a metal (for a review see \cite{mermin}). There is also a mixed case in which the bulk retains a gap, while there are gapless, topologically protected edge excitations; these are the famed topological insulators (for a review see \cite{hasan}). 

When the Coulomb interaction becomes so strong that it prevents charges to move without paying a large energy cost, these freeze and localize, even if band theory would predict a metal; this is a Mott insulator \cite{mott}. Finally, strong disorder can localize charges even in absence of interactions via the Anderson mechanism \cite{anderson}. Contrary to band insulators and Anderson insulators, which exist also for non-interacting particles, a Mott insulator is predicated on strong Coulomb interactions. When strong interactions co-exist with Anderson localization there is the possibility of a many-body localized phase, although this is established only in one space dimension (1D) \cite{altshuler}.  

What is often forgotten is that interactions can be strongly enhanced by the condensation of dual topological defects. Electromagnetic interactions are no exception. The condensation of magnetic monopoles (for a review see \cite{olive}), which are instantons in 2D and solitons in 3D, can lead to an electric potential which increases linearly with distance \cite{polyakov, polyakovbook}. As in the familiar case of 2D vortices (for a review see \cite{minnhagen}), this implies an infinite energy to separate a charge-hole pair and charges are confined. The electric confining interaction, however, is much stronger than the confinement of vortices in 2D, being linear rather than logarithmic, and therefore qualitatively analogous to the mechanism that binds quarks into hadrons (for a review see \cite{greensite}). As one can never observe quarks in isolation, in this phase one cannot observe charges in isolation at large distances and the resistance is thus strictly infinite, even at finite temperatures, up to a critical deconfinement temperature. Such materials are ``superinsulators" \cite{dst, vinokurnature, dtv}. 

Non-Abelian gauge theories with a compact group, like colour SU(3), are confining . The coupling constant runs to large values at large distances and non-perturbative effects lead to the formation of a string, a flux tube of chromoelectric field between colour charges (for a review see \cite{greensite}). In the case of the Abelian electromagnetic field, there are two possibilities. Either the gauge group is the non-compact group ${\mathbb R}$ or the compact group U(1), which is essentially a circle, as opposed to the real line. Nothing special happens in the former case. In the latter case, however, the model admits magnetic monopole solutions. These are solitons (non-trivial classical solutions of the field equations) in 3D and instantons in 2D. Instantons are non-trivial classical solutions of the Euclidean space-time field equations and represent tunnelling events in physical Minkowski space-time (for a review see \cite{coleman}). At strong coupling, magnetic monopoles render electromagnetism confining by generating an electric flux tube between charges and holes \cite{polyakov, polyakovbook}. These electric flux tubes are the dual of magnetic Abrikosov vortices in superconductors (for a review see \cite{tinkham}).  Superinsulators are thus simply dual superconductors. The resistance is infinite in the former and vanishes in the latter, electric fields are focused into thin flux tubes in the former, while the same happens to magnetic fields in the latter. 

There are two ways to formulate compact electrodynamics. Either one obtains the compact U(1) gauge group by spontaneous symmetry breaking from a larger compact non-Abelian symmetry or one formulates the theory on a lattice, with link-associated gauge fields periodic under finite shifts \cite{polyakovbook}. In the condensed matter setting it is this latter possibility that is realized. Superinsulators arise in systems with emergent granularity, where a superconducting condensate decomposes in separate islands, each with its own independent phase. Note that this does not imply physical granularity, the material can be perfectly homogenous. Of course, physically granular superconductors also fall in this category.  Granularity, emergent or physical, provides the natural mechanism for the manifestation of compact electrodynamics. 

Consider a 2D granular condensate, in which each island is characterized by a phase, such as it forms near the superconductor-to-insulator transition (SIT) \cite{goldman1, haviland, hebard, fisher1, fisher2, goldmanrev}. When phases are aligned, Cooper pairs can tunnel from one island to the other by the Josephson effect (for a review see \cite{tinkham}) and global superconductivity (to be distinguished from local superconductivity within the islands) sets in. When phases are disordered, though, vortices can form, represented by the circulation of the phases on the islands adjacent to a condensate void forming a multiple of $2\pi$. These are {\it not} Abrikosov vortices (for a review see \cite{tinkham}), though, they are point-like vortices with no dissipative core, analogous to the vortices of the XY model (for a review see \cite{minnhagen}), essentially point-like excitations associated with the ``dual lattice", the array of spaces between the superconducting grains. Accordingly, vortices can tunnel from one ``plaquette" to another by quantum phase flips on the grains, exactly as charges can tunnel from one grain to another via the Josephson effect. The only source of quantum dissipation associated with this vortex motion is the excitation of quasiparticles on the grains. For the relevant regime of high dielectric constant, however, the quasi-particle gap is much larger than the energy scale associated with tunnelling. Moreover, as we shall see, quasi-particles are simply absent due to confinement. Dissipation for vortex tunnelling is thus strongly suppressed at low temperatures and the physics of these 2D granular superconductors is that of a coupled system of dual excitations, charges and XY vortices. 

Charges and vortices cannot condense simultaneously since charge condensation creates a large cost for single vortices and vice versa. All three other possible phases, however are realized. Charge condensation corresponds to the original global superconductivity and vortex condensation, instead leads to superinsulation. However, contrary to what is sometimes claimed in the literature (see, e.g. \cite{phillipsrev1}), there can be an intermediate phase in which both charges and vortices are simultaneously out-of-condensate. This phase is a bosonic topological insulator in which metallic conduction is realized along topologically protected percolation edges, a phase often called a Bose metal (for a review see \cite{boserev}). Bose metal, superinsulator and the correct quantum phase structure of the SIT were first predicted in \cite{dst}. 

Contrary to Cooper pairs, which are Noether charges, vortices are topological excitations and they are thus not locally conserved, but only globally. They can tunnel in and out of the vacuum by instanton events. The vortex current is not locally conserved, as is the electric current, the only constraint is global neutrality of the instantons. In the Euclidean setting, a vortex at rest is a string of flux in the third direction, the one which becomes time in Minkowski signature. An instanton is an event where this flux flows out in Euclidean space and the vortex ceases to exist as such. This event must be compensated by an anti-instanton to ensure global neutrality. When vortices have tension, instanton anti-instanton pairs are bound into tight dipoles and are effectively suppressed; the vortex current becomes locally conserved as the electric one. When vortices condense, however, they become tensionless \cite{dst2}. The vortex string in the third direction attached to an instanton can be moved and deformed at will with no energy cost and becomes a Dirac string, while the instanton event becomes a magnetic monopole (for a review see \cite{olive}). When vortices proliferate, so do these magnetic monopole instantons \cite{dst2} and the plasma of such instantons has the effect of focusing the electric field between probe charges and holes into thin flux tubes, leading to a linearly confining potential \cite{polyakov, polyakovbook}. This causes a strictly infinite resistance (on large scales) and superinsulation is established. In this phase of matter, the effective electromagnetic action and the electromagnetic response are those of compact quantum electrodynamics (QED). This is what distinguishes a superinsulator from a Mott insulator, for which the effective electromagnetic action is non-compact. Essentially, superinsulation 
arises because: 1) a local treatment of dominant topological charge-vortex interactions requires a compact emergent gauge theory and 2) in this gauge theory a vortex condensate is not a gas of long loops but, rather, a monopole instanton plasma and this medium is confining. 

The SIT and the emergence of superinsulation rest ultimately on a competition between magnetic and electric interactions. In thin films (2D) it is the electric interaction which becomes logarithmic, and thus comparable to the interaction between vortices, due to a diverging dielectric constant . In bulk materials (3D) it is the magnetic interaction which becomes Coulombic, and thus comparable to the electric interaction, due to the presence of magnetic monopole solitons \cite{moncond}.  Both Bose metal \cite{van, clean1, clean2, marcus1, yang, kapitulnik, marcus2, marcus3} and superinsulating phases \cite{shahar, vinokurnature, nbtin, electrostatics, relaxation, bsi}  have by now been observed in numerous systems, including highly ordered ones, strongly disfavouring disorder-based explanations as the primary mechanism; topology and interactions are the dominant effects. This is further confirmed by a direct measurement of the confining linear potential in the superinsulating state \cite{relaxation}. 

In the whole of this review we shall use natural units for which $c=1$, $\hbar = 1$ and $\varepsilon_0 = 1$. We shall use boldface characters and latin indeces to indicate purely spatial locations and directions and Greek letters to denote spacetime directions, with the understanding that a summation is implied over equal Greek indices. The symbol $\epsilon_{\mu \alpha \nu}$ denotes the totally antisymmetric tensor. 

\section{Superconducting films near the SIT}

It is known since the 70s that the dielectric constant diverges at the metal-insulator transition \cite{efros1, efros2}. When the temperature is lowered, the metal may turn into a superconductor. At the corresponding quantum transition between this superconductor and an insulating state the dielectric constant also diverges \cite{baturina, bakkali}. We now explore the deep consequences of this divergence. 

Thin films are typically modelled as systems in which the matter degrees of freedom are constrained to live on a 2D plane, while electromagnetic fields can space out in 3D. However, when the dielectric constant becomes very large they choose dynamically not to do so. Let us consider a horizontal thin film of large dielectric constant $\varepsilon$. Above the film there is a medium of dielectric constant $\varepsilon_1$ and below the film there is a medium of dielectric constant $\varepsilon_2$. If the film is suspended in vacuum, we have $\varepsilon_1=\varepsilon_2 = 1$. If the film is deposited on a substrate, only $\varepsilon_1=1$. We are interested in the case in which $\varepsilon_1 \ll \varepsilon $ and $\varepsilon_2 \ll \varepsilon$. This problem has been considered and solved in \cite{chaplik, keldysh} (for a review see \cite{baturina}) and leads to the electric Keldish potential, which is a purely 2D logarithmic Coulomb potential up to an electric screening length $\lambda_s= d \varepsilon/(\varepsilon_1+\varepsilon_2)$, with $d$ the film thickness. 

In the symmetric case $\varepsilon_1=\varepsilon_2 =\varepsilon_{\rm out} \ll \varepsilon$, this result can be obtained with a much simpler calculation. From the boundary conditions
\begin{equation}
\varepsilon {\bf E}^{\perp}_{\rm in} = {\varepsilon_{\rm out}} {\bf E}^{\perp}_{\rm out} \ ,
\label{bc}
\end{equation} 
on the normal components of the electric field {\it in} the film and {\it out} of the film, we recognize that these normal components in the thin film are strongly suppressed. This implies that the potential varies negligibly across the thickness of the film. Any finite difference between the potentials at the two boundaries would require a transverse electric field E  whose electrostatic energy cost becomes prohibitive for large $\varepsilon$ and small thickness $d$. In the absence of free charges or external sources enforcing such a field, the system minimizes its energy by suppressing it, so that the potential takes the same value at the two boundaries. Therefore, the film can be effectively described by a two-dimensional electrostatic problem with periodic boundary conditions across its thickness

With periodic boundary conditions, the Coulomb potential $G$ among two charges $e$ in a film of thickness $d$ along the z-axis is determined by the equation
\begin{equation}
\nabla^2 G\left( |{\bf x}|, z, d \right)  = e^2 \delta^2 \left( {\bf x} \right) \sum_{n=-\infty}^{+\infty} \delta(z-nd) \ ,
\label{cpot1}
\end{equation}

The periodic delta function represents an infinite series of couples of mirror charges on the z-axis outside the film, whose effect is to mimic the periodic boundary conditions. We now recall that an electric dipole of length $\ell$ in a medium with dielectric constant $\varepsilon$ depends on the effective distance $\ell/\varepsilon$. Therefore, to maintain periodic boundary conditions with the same thickness $d$ but with a dielectric constant $\varepsilon$ inside the film we must place the mirror charges at distances $n \varepsilon d$,
\begin{equation}
\nabla^2 G\left( |{\bf x}|, z, d \right)  = e^2 \delta^2 \left( {\bf x} \right) \sum_{n=-\infty}^{+\infty} \delta(z-n\varepsilon d) \ .
\label{cpot2}
\end{equation}
The solution to this equation can be written in the form
\begin{equation}
G\left( |{\bf x}|, z, d \right) = {e^2\over 4\pi \varepsilon d} \sum_{n=-\infty}^{+\infty} \int_0^{\infty} dt {1\over \sqrt{1+t^2}} {\rm e}^{i {2\pi n\over \varepsilon d} \left( t|{\bf x}|-z \right)} \ . 
\label{cpot3}
\end{equation}
In the regime $d \ll |{\bf x}| \ll \varepsilon d$ we can keep only the first dipole $n=\pm 1$ and neglect $z$ in the ensuing cosine function, which gives
\begin{equation}
G\left( |{\bf x}|, d \right) =  {e^2\over 4\pi \varepsilon d} \int_0^{\infty} dt {1\over \sqrt{1+t^2}} {\rm cos} {2\pi \over \varepsilon d}  t|{\bf x}| =
{e^2\over 4\pi \varepsilon d} K_0 \left( 2\pi {|{\bf x}| \over \varepsilon d} \right) \ ,
\label{cpot4}
\end{equation}
where $K_0$ is the zeroth-order modified Bessel function of the second kind. As expected, both (square) charges and 2D distances are renormalized by a factor $1/\varepsilon$. This result becomes exact in the limit $d\to 0$, $\varepsilon \to \infty$ so that $\varepsilon d = \lambda_s$ is finite, in which case the Coulomb potential becomes a pure 2D one,
\begin{equation}
G\left( |{\bf x}|, d \right) = - {e^2\over 4\pi \lambda_s} {\rm ln} \left( {|{\bf x}| \over \lambda_s }\right) + {\rm constant} \ ,
\label{copt5}
\end{equation}
up to a screening length $\lambda_s$. Up to a numerical constant O(1) this is the Keldish result. The Keldish potential implies that, for very large dielectric constant $\varepsilon$, the electric screening length becomes typically larger than the lateral dimension $L$ of the sample, $L < \lambda_s$. In these films, electromagnetism is entirely 2D: the lines of electric field do not exit the film in the orthogonal direction. 

This has a significant consequence if the material in the film is a superconductor. The field theory governing a superconductor is the Abelian Higgs model or its non-relativistic version, the time-dependent Ginzburg-Landau model. In 2D these field theories are super-renormalizable because the electric coupling is dimensionful, essentially $e^2/d$, as we have shown above and the corresponding soft photons cause infrared divergences. These can be cured by an expansion in terms of ${\rm ln} (e^2 L/d )$ , where $L$ is the typical lateral dimension of the sample \cite{jackiw}, which means that the theory becomes non-perturbative in systems much larger than their thickness. As a consequence, when decreasing the thickness near the SIT, a superconducting film typically breaks up in many condensate islands of the typical lateral size $O(d)$, each characterized by its own phase, forming an emergent Josephson junction array (JJA) \cite{super2D, typeIII, typeIIIvort}. This by now paradigmatic condensate granularity has been verified for such materials as TiN \cite{sacepe1}, NbTiN \cite{nbtin} and InO \cite{granular, kapitulnik}. 

It becomes thus obvious that in these systems with granular condensate it is the compact version of electrodynamics which is realized. The emergent JJA provides the discrete ``lattice"
on which each ``node" (condensate grain) is characterized by its own independent phase which is an angular variable, thus realizing the compact gauge group U(1). Actually, such a system is essentially the paradigmatic realization of compact QED as described in Polyakov's original work \cite{polyakov, polyakovbook}. 

Note also that the irregularity of the emergent JJA is what is typically called ``disorder" in the some literature on the subject. This disorder is not relevant simply because topology and interactions cause much stronger versions of insulation before this disorder can even have an effect. 

\section{The Bose metal}
\begin{figure}[t!]
	\includegraphics[width=10cm]{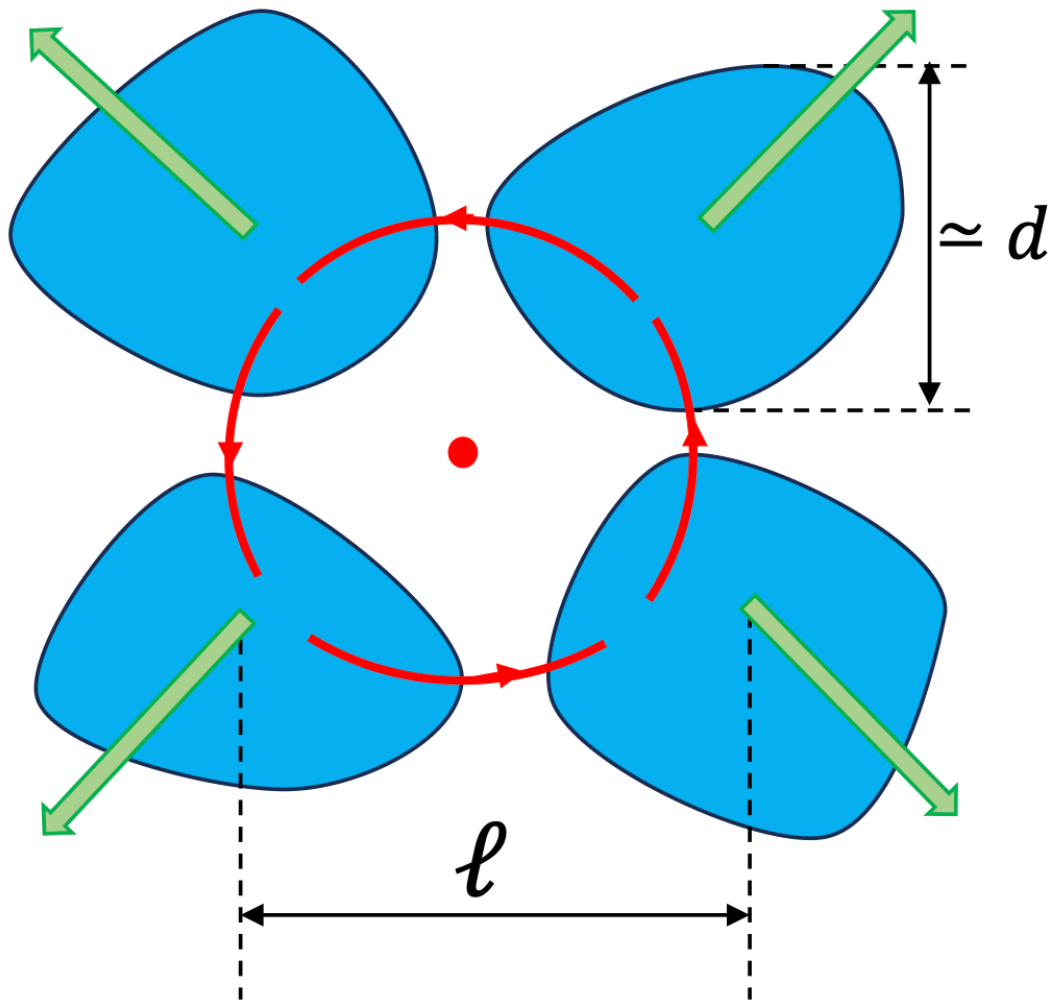}
	%\includegraphics[width=\textwidth]{phase_diagram.pdf}
	%\vspace{-1cm}
	\caption{An XY vortex in a superconducting film with emergent condensate granularity.}
	\label{fig:Fig.1}
\end{figure}

Let us consider a typical interspace between superconducting grains. The circulation of the phases on the adjacent grains forms an angle $2\pi n$, $n\in {\mathbb Z}$, as shown in Fig. \ref{fig:Fig.1} . If $n\ne 0$, there is a vortex at this interspace. Importantly, this is not an intra-grain Abrikosov vortex but an inter-grain XY vortex, which has no dissipative core. 

\begin{figure}[t!]
	\includegraphics[width=8cm]{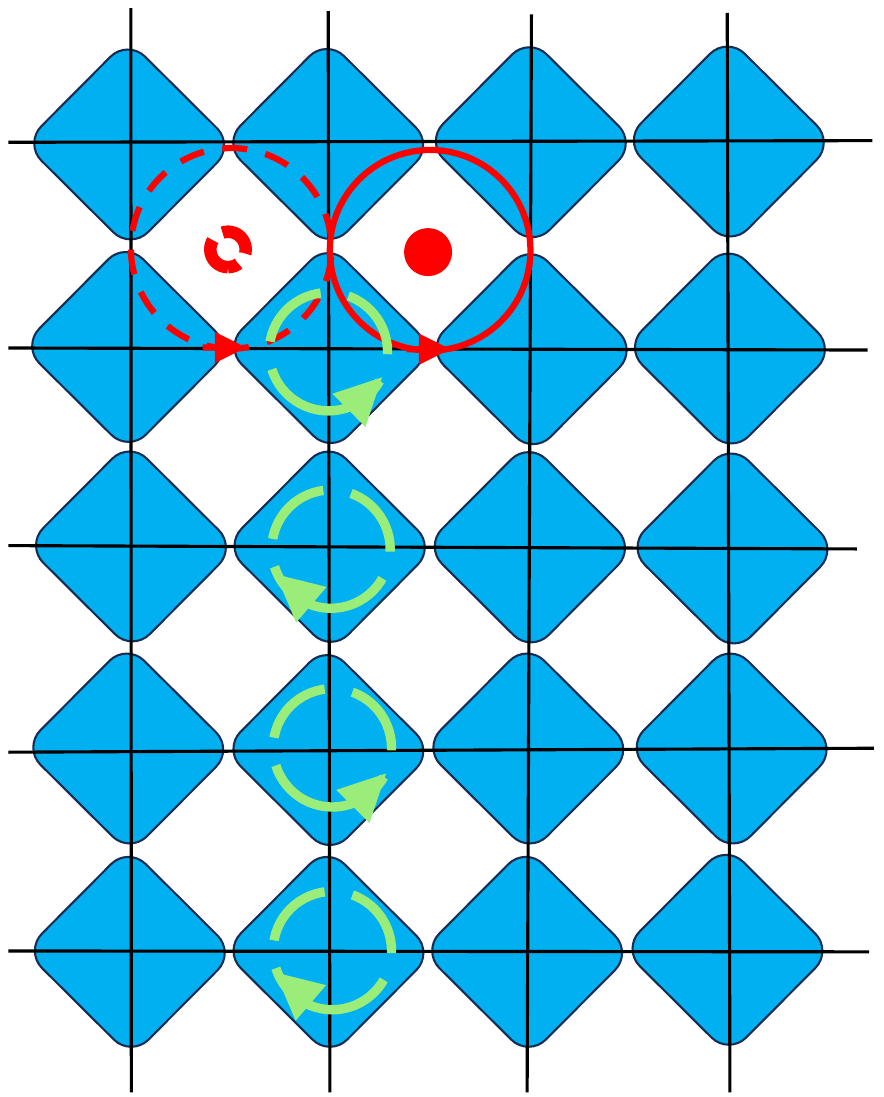}
	%\includegraphics[width=\textwidth]{phase_diagram.pdf}
	%\vspace{-1cm}
	\caption{Tunnelling of a single XY vortex as a line of phase slips.}
	\label{fig:Fig.2}
\end{figure}

Such a vortex can tunnel from one interspace to an adjacent one by a line of phase slips on the grains, as shown in Fig. \ref{fig:Fig.2}. 
Furthermore, since these vortices have no dissipative core, the only source of quantum dissipation for tunnelling could arise from the excitation of quasiparticles. However, for high dielectric constant, in which case the velocity $v \to 0$, the tunnelling is adiabatic and the energy scale associated with it is much smaller than the quasi-particle gap. Moreover, as we shall see, either vortices are frozen by topological interactions or quasi-particles are confined. Therefore, quantum dissipation for vortices is suppressed. The system reduces to a coupled model of charges (Cooper pairs) and vortices with dual behaviour. 

There are three types of interactions in such a system, charge-charge interactions, governed by an action $S_c$, vortex-vortex interactions, with action $S_v$ and charge-vortex interactions. These mixed interactions are topological, they are the ones that give rise to Aharonov-Bohm \cite{aharonovbohm} and Aharonov-Casher \cite{aharonovcasher} phases for closed trajectories of one type of excitations around the other one. As such, they are the dominant interactions in the whole model. However, being topological, they do not show up in the Hamiltonian. This is why approaches focusing on the Hamiltonian typically miss the most important ingredient; it is more appropriate to start from an action, instead. 

In the path integral formulation (see e.g. \cite{negele}), the Euclidean partition function for the system of charges and vortices was formulated in \cite{wu}, 
\begin{equation}
Z = \int {\cal D} Q {\cal D} M \ {\rm e}^{-S_c + S_v +i S_{\rm Top}} \ ,
\label{wuform}
\end{equation}
where ${\cal D} Q = {\cal D} x_1(s) \dots {\cal D} x_{N_c} (s)$, with an analogous expression for ${\cal D}M$, and $x_I(s)$ and $x_J(s)$ parametrize the charge and vortex trajectories $C_I$ and $C_J$, respectively. The topological contribution to the action is
\begin{eqnarray}
S_{\rm top} &&=  2\pi  \sum_{I,J} Q_I M_J \Phi \left( C_I, C_J \right) \ ,
\nonumber \\
\Phi \left( C_I, C_J \right) &&= {1\over 4\pi} \int_0^1 ds \int_0^1 dt \ {dx_I^{\mu} \over ds} \epsilon_{\mu \alpha \nu}
{(x_I -x_J)^{\alpha} \over |x_I-x_J|^3} {dx_J^{\nu} \over dt} \ .
\label{gauss}
\end{eqnarray} 
For closed curves $C_I$ and $C_J$, the quantity $\Phi \left( C_I, C_J \right)$ is the integer Gauss linking number of the two trajectories (see, e.g. \cite{kaufmann}). If the quantum numbers are such that $Q_I M_J = {\rm integer} $ for all $I, J$, the phases are trivial for closed trajectories. This means that the overall wave function of the many-body system acquires a factor (-1) upon the interchange of a charge and a vortex. This is the concept of relative statistics: charges $Ne$ and vortices $2\pi/Ne$ are relative fermions and correspondingly statistically repel each other. 

The relative statistics constraints must be taken into account for a correct description of the charge-vortex system. One way to go is the Wu formulation above. 
Unfortunately, however, the phases in (\ref{gauss}) are non-local, and thus unsuited for the machinery of quantum field theory. This can be remedied by the introduction of emergent gauge fields \cite{wilczek}, in the same way as gauge fields can realize the usual fermionic statistics in 2D \cite{wilczekbook}. Contrary to this well-known case, though, here there are two types of excitations with relative statistics and, thus, two gauge fields are necessary.  

To see how this works let us first introduce charge and vortex current fields
\begin{eqnarray}
Q_{\mu} &&= \sum_I Q_I \int ds \ {dx_I^{\mu} \over ds} \ \delta^3 \left( x -x_I (s) \right) \ ,
\nonumber \\
M_{\mu} &&= \sum_J M_J \int dt \ {dx_J^{\mu} \over dt} \ \delta^3 \left( x -x_J(t) \right) \ .
\label{pointsources}
\end{eqnarray}
Then we can represent the relative statistics phases as 
\begin{equation}
S_{\rm top}  =  \ 2\pi   \int d^3x \  Q_{\mu} \epsilon_{\mu \alpha \nu} {\partial_{\alpha} \over \nabla^2} M_{\nu} \ ,
\label{effec3}
\end{equation}
where $\epsilon^{\mu \alpha \nu}$ denotes the totally antisymmetric tensor and 
\begin{equation}
{1\over -\nabla^2} \delta^3 (x) = {1\over 4\pi} {1\over |x|} \ ,
\label{3dG}
\end{equation} 
is the 3D Poisson Green function. 

While this is still non-local, a local formulation can be obtained \cite{wilczek} by coupling the charge and vortex currents to two emergent gauge fields $a_{\mu}$ and $b_{\mu}$ with a mixed Chern-Simons interaction \cite{jackiw2},
\begin{eqnarray}
Z_{\rm top} &&= \int {\cal D}Q_\mu {\cal D} M_\mu  \int {\cal D} a_{\mu} {\cal D} b_{\mu} {\rm e}^{-S_{\rm top}} \ ,
\nonumber \\
S_{\rm top} &&= \int d^3 x {i\over 2\pi} a_{\mu} \epsilon^{\mu \alpha \nu} \partial_{\alpha} b_{\nu} + i a_{\mu} Q_{\mu} + i b_{\mu} M_{\mu} \ .
\label{wilc}
\end{eqnarray}
The original Wu formulation is obtained by integrating over the two emergent gauge fields. This action is both parity- and time-reversal-invariant since $a_{\mu }$ is a vector field, as usual, but $b_{\mu}$, coupling to the vortex current, is a pseudovector gauge field. 

In modern parlance, this action, first derived as a description of a concrete condensed matter system in \cite{dst}, describes a bosonic topological insulator \cite{bti1, bti2, bti3}. Indeed, the simplest bosonic topological insulator is nothing else than a two-component liquid of relative fermions. The equations of motion are derived from the Minkowksi space-time version of (\ref{wilc}), 
\begin{eqnarray}
f^{\mu} = {1\over 2} \epsilon^{\mu \alpha \nu}  f_{\alpha \nu} = \epsilon^{\mu \alpha \nu} \partial_{\alpha} b_{\nu} =2\pi Q^{\mu} \ ,
\nonumber \\
g^{\mu} = {1\over 2} \epsilon^{\mu \alpha \nu}  g_{\alpha \nu} = \epsilon^{\mu \alpha \nu} \partial_{\alpha} a_{\nu} =2\pi M^{\mu} \ .
\label{fields}
\end{eqnarray}
and imply that the dual field strengths $f^{\mu}$ and $g^{\mu}$ associated with the two emergent gauge fields represent ($2\pi$ times) the conserved charge and vortex (number) currents, respectively.

Since the action is of first-order in time derivatives, the canonical commutation relations can be simply read off the (Minkoswki space-time) action after setting the Lagrange multipliers $a_0=0$ and $b_0=0$, the Weyl gauge for the gauge theory \cite{djt}, 
\begin{equation}
\left[ a_i({\bf x}), b_j ({\bf y}) \right] = -i \epsilon^{ij} \delta^2 \left({\bf x} -{\bf y} \right)\ ,
\label{comm1}
\end{equation}
where boldface characters denote purely spatial coordinates and all commutators among $a$ and $b$ gauge fields themselves vanish. Using (\ref{fields}) this implies that
\begin{equation}
\left[ Q({\bf x}), M({\bf y}) \right] = 0 \ .
\label{nocomm}
\end{equation}
This vanishing commutator implies that there is no reason why charges and vortices cannot be both simultaneously out of condensate. 

When this happens, statistical frustration immediately leads to a frozen state. Similarly to the quantum Hall effect, where a longitudinal current creates a transverse voltage, here a longitudinal vortex motion creates a transverse voltage for charges and viceversa. Longitudinal motion in one sector creates a transverse voltage in the dual sector. Therefore, any motion leads to an increase in energy and the system is effectively frozen. The pure Chern-Simons theory (\ref{wilc}) represents the limit in which the energy gap for motion is infinite and the system settles in a ``mutual Hall liquid" \cite{bm, boserev}. It is important to stress that this frozen state is not due to disorder, as often claimed \cite{phillips3, phillips4, phillipsrev1, kapitulnik} but, rather to the topological effect of mutual statistical frustration \cite{boserev, nodisorder}. 

While the bulk is frozen, motion, and thus charge transport, can occur near the edges of the system, where the frustration effect is imperfect. As in the case of quantum Hall liquids, the system supports gapless edge excitations \cite{wen}, in this case electric ones propagating along the edges and magnetic ones crossing the edges. From the point of view of the edges, these latter correspond to 1D quantum phase slips \cite{qps1, qps2}, resulting in a finite metallic resistance \cite{bm, boserev}. 

To obtain the edge physics let us minimally couple the Chern-Simons action to the real electromagnetic field $A_{\mu}$ (in Minkowski space-time),
\begin{equation}
S={1\over 2\pi} \int d^3x \ a_{\mu} \epsilon^{\mu \alpha \nu} \partial_{\alpha} b_{\nu} - {2e\over 2\pi} \int d^3x \ A_{\mu} \epsilon^{\mu \alpha \nu} \partial_{\alpha} b_{\nu}. 
\label{mincou}
\end{equation}
This action is not gauge invariant under transformations that do not vanish on the egdes and one must add edge degrees of freedom so that the gauge invariance of the complete model is restored, as for the quantum Hall states \cite{wen}. To do so we consider the Weyl gauge, $a_0=0$, $b_0=0$, and we set the spatial field components as pure gauges, $a_i=\partial_i \lambda$, $b_i=\partial_i \chi$. To restore full gauge invariance, these fields have to be promoted to new scalar degrees of freedom on the edges, with an action
\begin{equation}
S_{\rm e} = {1 \over 4\pi} \int d^2 x \ \left( \partial_0 \lambda \partial_s \chi + \partial_0 \chi \partial_s \lambda \right) +
{2e \over 2\pi} \int d^2 x \  \left( A_0  \partial_s \chi - A_s \partial_0 \chi \right) ,
\label{edgeac}
\end{equation}
where $s$ denotes the space coordinate along the edge. We now introduce two new edge fields by the relations $\lambda = \xi + \eta$ and $\chi= \xi-\eta$ and we set $A_s=0$ to focus on a configuration with a constant edge electric field $E=\partial_s A_0$,
\begin{equation}
S_{\rm e} = {1 \over 2\pi} \int d^2 x \ \left( \partial_0 \xi \partial_s \xi - \partial_0 \eta \partial_s \eta \right) +
2e  \int d^2 x \   A_0  \left( {1\over 2\pi} \partial_s \chi \right) .
\label{edgeacmod}
\end{equation}
From here we recognize $\rho = \partial_s \chi /2\pi$ as the one-dimensional charge density of edge excitations. 

As in the quantum Hall case \cite{wen}, we must add a non-universal dynamics for the edge modes, with Hamiltonian
\begin{equation}
H_{\rm e} = {1 \over 2\pi} \int ds \left[ v_b \left( \partial_s \xi \right)^2 + v_b \left( \partial_s \eta \right)^2\right] \ ,
\label{four}
\end{equation}
where $v_b$ is the velocity of propagation of the edge modes along the boundary. The total edge action then becomes
\begin{eqnarray}
S_{\rm e} &&= {1 \over 2\pi} \int d^2 x \ \left[ \left(\partial_0-v_b\partial_s\right) \xi \partial_s \xi  - \left(\partial_0+v_b \partial_s \right)\eta \partial_s \eta \right]\nonumber \\
&&+ 2e  \int d^2 x \  A_0  \left( {1\over 2\pi} \partial_s \chi \right) \ ,
\label{edgeacfull}
\end{eqnarray}
with the corresponding equations of motion
\begin{eqnarray} 
\left( \partial_0 -v_b \partial_s \right) \partial_s \xi &&= - e E \ ,
\nonumber \\
\left( \partial_0 +v_b \partial_s \right) \partial_s \eta &&= - e E \ ,
\label{eq1}
\end{eqnarray}
or, in the original fields,
\begin{eqnarray} 
\partial_0 \partial_s \lambda -v_b \partial_s^2 \chi &&= - 2e E \ ,
\nonumber \\
\partial_0 \partial_s \chi - v_b \partial_s^2\lambda &&= 0\ .
\label{eq2}
\end{eqnarray}
Using the edge current $I = 2e v_b \rho$, we can reformulate the first of these equations as
\begin{equation}
\partial_s I = {4e^2 \over 2\pi} E + {2e\over 2\pi} \partial_0 \partial_s \lambda \ .
\label{fincur}
\end{equation}
Remembering that $(1/2\pi) \epsilon^{\mu \alpha \nu} \partial_{\alpha} a_{\nu}$ is the vortex number current, we note that $(1/2\pi) \partial_0 \partial_s \lambda$ is the number of vortices crossing the edge in a given orthogonal direction. On the right-hand side of (\ref{fincur}) we recognize the two mechanisms by which charges are accelerated/decelerated along the edge, an external electric field and quantum phase slips. 

To compute the resistance of the edges we will time-average the above equation for the current, which gives the steady-state value, 
\begin{equation}
\langle I \rangle = {4e^2\over 2\pi} \left( V + {1\over 2e} \langle \partial_0 \lambda \rangle \right) \ ,
\label{add1}
\end{equation}
with $V$ the applied voltage, $E= \partial_s V$. This shows that quantum phase slips act as an additional, effective voltage on the edge. Defining this effective voltage as
\begin{equation}
V_{\rm QPS} = {1\over 2e} \langle \partial_0 \lambda \rangle = \kappa V \ ,
\label{add2}
\end{equation}
we can express the induced current as
\begin{equation}
\langle I \rangle = {4e^2\over 2\pi} (1+\kappa) V \ ,
\label{add3}
\end{equation} 
which gives the sheet resistance
\begin{equation}
R_{\square} = {2\pi \over g (2e)^2} = {R_{\rm Q} \over g} 
\label{qr}
\end{equation}
where $R_{\rm Q} = 2\pi /4e^2$ is the resistance quantum and $g=(1+\kappa)$. This can be viewed as the resistance quantum for an effective charge $2e_{\rm eff} =2e \sqrt{g}$ renormalized by quantum phase slips. 

When interactions with quantum phase slips are totally random we have $\kappa =0$ since the time-average of $\partial_0 \lambda$ vanishes. In this case the edge mode sheet resistance is exactly the resistance quantum. This is the universal metallic point originally discovered in \cite{fisher1, fisher2}. If there is an excess of accelerating quantum phase slips, the sheet resistance is smaller than the quantum. On the contrary, if there is an excess of slowing quantum phase slips, the resulting sheet resistance is larger than the quantum. These two possibilities are realized in the broadening of the universal metallic point to a full fledged Bose metal phase originally predicted in \cite{dst} and then rediscovered two years later in \cite{das} (for a review see \cite{boserev}). It is important to stress that edge transport is topologically protected from localization and disorder-induced dissipation \cite{bti1, bti2, bti3}. Quantum phase slips are the only dissipation mechanism. Moreover, edges need not be exclusively the external boundaries of the sample. There can be as well internal edges forming a percolation network, exactly as in the quantum Hall case \cite{cc}. 

The Bose metal phase has been first detected in a Josephson junction array (for a review see \cite{jjarev}) the same year it was theoretically predicted \cite{van}. It has by now been measured in a wide variety of materials, natural and synthetic alike \cite{kapitulnik, bm, clean1, clean2, marcus1, marcus2, marcus3}. In particular, it is now established experimentally that vortices are out of condensate and mostly frozen in the Bose metal state, while a small fraction remain mobile and cause the residual metallic sheet resistance saturation \cite{marcus2}. Note that the observation of the Bose metal in ordered synthetic materials conclusively dispels the notion that this phase of matter is due to disorder. 

\section{The quantum phase structure of thin superconducting films} 

Having established that the infrared (IR) dominant contribution in a local theory of interacting charges and vortices is a Chern-Simons term mixing two emergent gauge fields coupled to the corresponding charge and vortex currents, we proceed according to the standard principles of effective field theory. We therefore supplement the action with all possible relevant or marginal operators compatible with the symmetries of the system, here, in particular, gauge invariance. The next-order gauge invariant terms in such an expansion are the Maxwell terms for the two gauge fields, which incorporate, of course, the effects of the material
\begin{equation}
\Delta S= \sum_{t,{\bf x}} {\ell_0 \ell^2} \left( 
{v\over 2e^2_v \mu} {\cal B}_b^2+ {v \varepsilon \over 2e^2_v} \vec{{\cal E}}_b^2
+{v \over 2e^2_q\mu} {\cal B}_a^2 + {v \varepsilon \over 2e^2_q} \vec{{\cal E}}_a^2 \right) \ ,
\label{nonrelac1}
\end{equation}
where we have formulated the model on a lattice with coordinates $\{ t, x_1, x_2 \}$ and spacing $\ell$. Since charges and vortices are quantized in terms of integers, we have to consider a (Euclidean) time variable which lies also on a lattice. To maintain relativistic invariance in the long wavelength limit while describing the interior of a material we choose the lattice spacing $\ell_0 = \ell / v$, where $v = 1/\sqrt{\varepsilon \mu}$ is the speed of propagation in the material (in units of $c$). ${\cal B}_{a,b}$ and $\vec{{\cal E}}_{a,b}$ are the magnetic and electric fields associated with the two emergent gauge fields, expressed, as usual, in terms of scalar and vector potentials $(\phi, {\bf a})$ with a similar expression for the other gauge field. 

There are four phenomenological parameters in this action. The couplings $e_{\mathrm v}^2$  and $e_{\mathrm q}^2$ distinguish between the electric and magnetic sectors of the theory and have canonical dimension [1/length]. They encode, up to numerical factors, the electric coupling constant derived above and the 2D superfluid stiffness of the XY vortices,
\begin{eqnarray}
e_{\mathrm q}^2 = {\rm {\cal O}} \left( {e^2\over 2\pi d} \right) \ ,
\nonumber \\
e_{\mathrm v}^2 = {\rm  {\cal O}} \left( {\pi d \over e^2 \lambda_{\rm L}^2}\right)  \ ,
\label{couplings}
\end{eqnarray}
where $\lambda_{\rm L}$ is the London penetration depth of the grains (see e.g. \cite{tinkham}). The two material parameters $\varepsilon$ and $\mu$ can be traded for the speed of propagation $v=1/\sqrt{\varepsilon \mu}$ and the ratio $\sqrt{\varepsilon /\mu}= 1/v\mu$. In the following we consider only materials for which $\mu=1$, which leaves only three parameters setting the dynamics of the gauge fields,
\begin{equation}
\Delta S= \sum_{t,{\bf x}} {v\ell_0 \ell^2} \left( 
{1\over 2e^2_v } {\cal B}_b^2+ {1\over 2e^2_v v^2} \vec{{\cal E}}_b^2
+{1 \over 2e^2_q} {\cal B}_a^2 + {1 \over 2e^2_q v^2} \vec{{\cal E}}_a^2 \right) \ .
\label{nonrelac2}
\end{equation}

Readers familiar with quantum field theory might object at this point that, in 2D, Maxwell terms are neither relevant nor marginal but, rather, power-counting irrelevant, since their coupling have dimension 1/[length]. When sending these coupling to $\infty$ to recover large-scale behaviour, though, one must specify how they diverge with respect to each other. In other words their ratio is a dimensionless marginal parameter that has a substantial role in determining the phase structure. This is why one must keep the Maxwell terms even if they are power-counting irrelevant, each considered alone. Indeed the two couplings $e_q$ and $e_v$ can be traded for a mass 
\begin{equation}
m = {e_{\mathrm q} e_{\mathrm v} \over 2\pi v} = {\cal O}\left( {1\over v \lambda_{\mathrm L}} \right) \ ,
\label{tom}
\end{equation} 
and one dimensionless parameter 
\begin{equation}
g={e_{\mathrm v}\over e_{\mathrm q}} = {\cal O}\left( {d\over e^2 \lambda_{\mathrm L}}\right) \ .
\label{gpar}
\end{equation}

We now introduce a new coordinate $x_0=vt$ with spacing $\ell$, the usual gauge fields $a_0=\phi / v$ and $b_0$ (defined the same way) and their dual field strengths $g_{\mu}$ and $f_{\mu}$. In terms of these variables the total action takes the form
\begin{equation} 
S=\sum_x \ell^3 \left( {1\over 2\pi} a_{\mu}k_{\mu \nu}b_{\nu} +{1\over 2e^2_v } f_{\mu}f_{\mu} +{1 \over 2e^2_q} g_{\mu}g_{\mu} \right) + i\ell a_{\mu} Q_{\mu} +i\ell b_{\mu} M_{\mu}\ ,
\label{totac}
\end{equation}
where $k_{\mu \nu}$ is the lattice version of the Chern-Simons operator (see Appendix A) and the dual field strengths on the lattice are defined as
\begin{eqnarray}
f^{\mu} &&= k_{\mu \nu} b_{\nu} =2\pi j_{\mu} \ ,
\nonumber \\
g^{\mu} && = k_{\mu \nu} a_{\nu} =2\pi \phi_{\mu} \ .
\label{fieldslatt}
\end{eqnarray}
The quantities $Q_{\mu}$ and $M_{\mu}$ are integer link variables forming closed loops which represent the Cooper pair and the vortex currents, respectively. 
The total charge current and vortex currents $j_{\mu}$ and $\phi_{\mu}$ now comprise both the Cooper pairs $Q_{\mu}$ and vortices $M_{\mu}$ and the backreactions of the now dynamical gauge fields. $f_0$ and $g_0$ encode the emergent ``magnetic" fields associated with $b_{\mu}$ and $a_{\mu}$, respectively, ${\bf f}$ and ${\bf g}$ the corresponding emergent ``electric" fields. The term ${\bf f}$ represents a kinetic term for charges while the corresponding term ${\bf g}$ encodes the dual kinetic term for vortices. As was pointed out in \cite{fazio}, such a kinetic term for vortices is induced by integrating over charge fluctuations. It is a fundamental principle of field theory that what is induced radiatively must be included in the tree level action. This is why early phase-only treatments are incomplete: omitting the topological mutual statistics interaction of charges and vortices leads to also missing the important role of a kinetic term for vortices and the requirement of a gauge theory rather than an XY-type model \cite{dst}. The experimental observation of the charge BKT (VFT) transition \cite{vinokurnature, nbtin, bsi} and the direct measurement of a linear confining potential \cite{relaxation} bear out this point and raise serious doubts that a Mott insulator provides the correct description of the insulating state at the SIT.

The model (\ref{totac}) is known as mixed (or mutual) Maxwell-Chern-Simons model \cite{dst}. In absence of external sources it describes two topologically massive \cite{jackiw} gauge field with mass (\ref{tom}) and dispersion relation
\begin{equation}
E = \sqrt{m^2 v^4 + v^2 p^2} \ .
\label{dis}
\end{equation}
When minimally coupled to $Q_{\mu}$ and $M_{\mu}$ it encodes the most general model of interacting charges and vortices in a 2D granular medium with $\mu=1$ up to two derivatives in the fields.

We now integrate over the gauge fields to obtain the induced action for the Cooper pairs and vortices. This gives,
\begin{eqnarray} 
S_{\rm ind} &&= \sum_{x,y} e_q^2 \ell Q_{\mu}(x) \ G(m\ell v; x,y) \ Q_{\mu} (y) + e_v^2 \ell M_{\mu}(x) \ G(m\ell v; x,y) \ M_{\mu} (y) 
\nonumber \\
&& + i 2\pi v^2 m^2 Q_{\mu}(x) {G(m\ell v; x,y)\over \Delta_3}  \ell k_{\mu \nu} M_{\nu}(y)\ ,
\label{indac}
\end{eqnarray}
where $\Delta_3 $ is the 3D Laplacian (see Appendix A) and $G(m\ell v; x)$ the Green function defined by
\begin{equation} 
{\ell^2\over v^2} \left( m^2 v^4 -d_t \hat d_t -v^2 \nabla^2 \right) G(m\ell v; x)= \ell^2 \left( m^2 v^2 -\Delta_3 \right) G(m\ell v; x) 
= \delta_{x,0} \ ,
\label{green}
\end{equation}
with $\nabla^2$ the 2D Laplacian and $d_t$ and $\hat d_t$ the forward and backward derivatives in the (Euclidean) time direction (see Appendix A). The last term is the lattice version of the Gauss link number. For $m\ell v \gtrsim 1$, when the dimension $1/m$ of charges and vortices is smaller than the lattice spacing, it reduces again to an integer. This can be easily seen by formulating the purely transverse $M_{\mu} = \ell k_{\mu \nu} S_{\nu}$, with $d_{\mu} S_{\mu}=0$ to maintain the correct number of degrees of freedom, and using the square of the $k$ operator to obtain a 3D Laplacian (see Appendix A) which cancels the one in the denominator in (\ref{indac}). For $m\ell v \gtrsim 1$ the remaining expression reduces to $i 2\pi \times {\rm integer}$ whose exponential drops out of the partition function. This takes the form of a statistical mechanics model with ``Hamiltonian" given by
\begin{equation} 
S_{\rm ind} = \sum_{x} 2\pi v m\ell G(m\ell v) \left( {1\over g} Q_{\mu}^2 + g M_{\mu}^2 \right) \ .
\label{ha}
\end{equation}
where we have approximated the short-range kernel (\ref{green}) by its diagonal element $G(m\ell v)$. 

\begin{figure}[t!]
	\includegraphics[width=12cm]{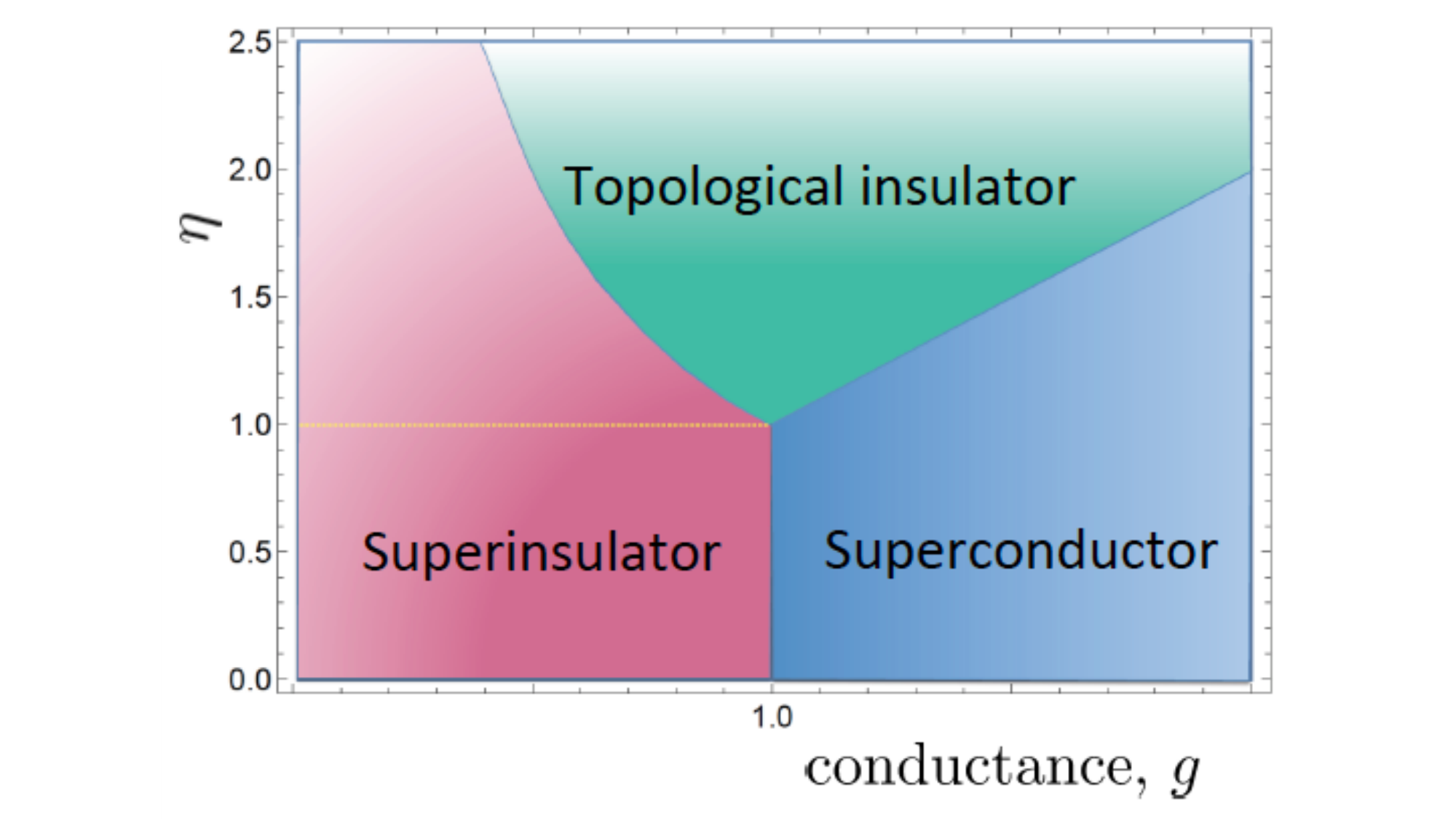}
	%\includegraphics[width=\textwidth]{phase_diagram.pdf}
	%\vspace{-1cm}
	\caption{The quantum phase diagram of granular superconducting films near the SIT.}
	\label{fig:Fig.3}
\end{figure}

Let us now consider a closed string made of $N$ links carrying integer electric and magnetic quantum numbers $Q$ and $M$, respectively. The total ``energy" (\ref{ha}) of such a string is proportional to its length $N$. The ``entropy" $\mu N$ of such a string is also proportional to its length, $\mu N$, with $\mu \approx {\rm ln} 5$, since, at each step, the non-backtracking string can choose among 5 possible direction on how to continue. This gives a string free energy
\begin{equation} 
F = 2\pi v m\ell G(m\ell v) \left( {1\over g} Q^2 + g M^2 -{1\over \eta} \right) N\ ,
\label{fe}
\end{equation}
where we have introduced the new dimensionless parameter
\begin{equation} \eta = {2\pi vm\ell G(m\ell v) \over \mu} \ .
\label{eta}
\end{equation}
The ground state of the quantum model is found by minimizing this free energy as a function of $N$. When the energy term in (\ref{fe}) dominates, the free energy is positive and consequently minimized by short loop configurations. When, instead, the entropy dominates, the free energy becomes negative and is minimized by large loops. The condition for condensation of long strings with integer quantum numbers  $Q$ and $M$ is given by
\begin{equation}
\eta {e_{\mathrm q}\over e_{\mathrm v}} Q^2 + \eta {e_{\mathrm v}\over e_{\mathrm q}} M^2 < 1 \ .
\label{ellipse}
\end{equation}
If two or more condensations are allowed, one has to choose the one with the lowest free energy. 
This condition describes the interior of an ellipse with the semiaxes 
\begin{eqnarray}
r_{\rs Q} = \sqrt{{e_{\mathrm v}\over e_{\mathrm q}} {1\over \eta}} \ ,
\nonumber \\
r_{\rs M}= \sqrt{{e_{\mathrm q}\over e_{\mathrm v}} {1\over \eta}} \ ,
\label{semiaxes}
\end{eqnarray}
on a square lattice of integer electric and magnetic charges. The quantum phase diagram is consequently found by establishing which integer charges lie within the ellipse when the semi-axes are varied,
\begin{eqnarray}
\eta &&< 1 \to 
\begin{cases}
{g}>1\ , {\rm electric \ condensation \ = \ superconductor} \ ,\\
{g}<1\ , {\rm magnetic \ condensation \ = \ superinsulator} \ ,\\
\end{cases}
\nonumber \\
\eta &&> 1 \to
\begin{cases}
{g} > {\eta}
\ , {\rm electric \ condensation \ = \ superconductor} \ ,\\
{\eta}
>{g}>{1\over \eta} \ , {\rm no \ condensation \ = \ Bose \ metal} \ ,\\
{g} < {1\over \eta} \ , {\rm magnetic \ condensation \ = \ superinsulator }\ ,\\
\end{cases}
\nonumber
\end{eqnarray}
and is shown in Fig. \ref{fig:Fig.3} as a function of the two relevant quantum couplings $g$ and $\eta$. For $\eta < 1$, there is a coexistence region of electric and magnetic charges when $g$ lies between $\eta $ and $1/\eta$. This indicates that the direct transition from a superinsulator to a superconductor is actually a ``quantum first-order transition" with metastable coexistence of phases in a region around it. Exactly this phase structure has been recently confirmed in these types of materials \cite{kapitulnik}. 

\section{The role of topological gauge fields in superconductivity}

In standard s-wave superconductors, the gauge field A enters the action through the Stückelberg combination $(A-d\phi)$, where $\phi$ is the phase of the order parameter. In homogeneous systems on 
 ${\mathbb R^D}$ , this combination depends only on the transverse component of A, giving rise to the Anderson–Higgs mechanism. As pointed out in \cite{hansson} and \cite{typeIII, higgsless}, the situation changes in the presence of non-trivial manifold topology with non-contractible loops. In this case, the Stückelberg combination acquires additional contributions from harmonic modes winding around these loops. These global modes are governed by a topological gauge field theory. When vortices and charges carry non-trivial winding, such as in ${\mathbb Z}_2$ superconductors, the ground state acquires a  topological degeneracy.

While this is mathematically interesting, because it implies that superconductors are topological states of matter, its practical relevance is limited in conventional settings, since typical superconductors are realized on simply connected geometries where such global modes do not lead to observable degeneracies. However, non-trivial topology of the underlying manifold is not the only way to generate non-contractible loops and non-trivial windings. In granular superconductors, as discussed above, phase windings arise around emergent ``plaquettes" defined by the network structure. These windings correspond precisely to the XY vortices of the system and are unrelated to global manifold topology. Unlike the sparse global modes in topologically non-trivial manifolds, these vortex degrees of freedom are abundant and play a central role in determining the phase structure. As first recognized in \cite{dst}, non-trivial winding modes in superconductors are described by a topological gauge theory. This theory governs {\it global} superconductivity across the system, in contrast to the Landau–Ginzburg description, which captures {\it local} superconductivity within individual grains. Global superconductivity is now referred to as type-III superconductivity \cite{typeIII, typeIIIvort}. Type-III superconductivity is not destroyed by Cooper pair breaking, but by the liberation of winding modes. When vortices proliferate freely, the system enters a bosonic topological insulator phase; when they condense, it transitions into a superinsulating phase, as we show below. Single-charge quasiparticles remain gapped at energies higher than the topological gap associated with windings and are therefore irrelevant for the global superconducting properties. Correspondingly, no ground state degeneracy arises in this case.

\section{Confinement and the charge BKT transition}
In this section we characterize the nature of the superinsulating phase, realized when vortices condense. To this end we couple the total electric current in (\ref{fieldslatt}) to the electromagentic field $A_{\mu}$ and we derive the induced electromagnetic action in this phase,
\begin{equation}
S \to S+ i\sum_x \ell^3 A_{\mu} j_{\mu} = S+ i\sum_x {\ell \over 2\pi} b_{\mu}  \left(\ell^2 F_{\mu}\right) \ ,
\label{elcou}
\end{equation}
where $F_{\mu} = \hat k_{\mu \nu}$ is the dual electromagnetic field strength. Comparing with (\ref{totac}) we realize that we can compute the induced electromagnetic action by simply replacing $M_{\mu} \to M_{\mu} +(\ell^2/2\pi) F_{\mu}$ in (\ref{ha}). 

In original anisotropic variables, the induced action at finite temperature $T= 1/\beta \ell_0$ can be formulated as
\begin{equation}
S_{\rm ind} =\sum_{t=1}^\beta \sum_{{\bf x}}\ell_0 \ell^2 \ g {\mu \eta \ell \over 4\pi^2} \left[ v \left( F_0 +{2\pi \over \ell^2} M_0^2 \right)^2 +{1\over v} \left( F_i +{2\pi \over \ell_0 \ell} M_i^2 \right)^2 \right] \ ,
\label{ind1}
\end{equation}
where $\beta$ is the number of links in the Euclidean time direction and we have neglected the charge fluctuations $Q_{\mu}$ which are suppressed in this phase. Of course, the diagonal element of the kernel (\ref{green}) in $\eta$ has to replaced by its finite-temperature version too. The zero-temperature diagonal element 
$G(m\ell v)$ is given by
\begin{equation}
G(m\ell v)= \int_{-\pi}^{+\pi} {d^3k\over (2\pi)^3} \ {1\over (m\ell v)^2 + \sum_{i=0}^2 4\  {\rm sin} \left( {k^i \over 2} \right)^2} \ ,
\label{zerogreen}
\end{equation}
while its finite-temperature counterpart becomes
\begin{equation}
G(m\ell v, \beta) = \sum_{k=-\beta}^{+\beta}{\pi \over \beta} \int_{-\pi}^{+\pi} {dk^1 dk^2 \over (2\pi)^3} \ {1\over (m\ell v)^2 + 4 \ {\rm sin} \left( \pi k\over 2\beta \right)^2 + \sum_{i=1}^2 4\  {\rm sin} \left( {k^i \over 2} \right)^2} \ .
\label{betagreen}
\end{equation}
By numerical computation one can verify for values over various orders of magnitude around $m\ell v=1$ that the ratio $R(\beta) = G(m\ell v, \beta) / G(m\ell v)$ depends only on temperature and increases when the temperature is raised ($\beta$ is lowered). As a consequence, $\eta (\beta)$ also increases at higher temperatures and the ellipse semiaxes in (\ref{semiaxes}) shrink. When the temperature is raised sufficiently, quantum numbers that were inside the ellipse find themselves outside and the corresponding condensates cease to exist. 

We shall now consider the neighbourhood of the SIT, where $v \ll 1$ and we neglect thus magnetic with respect to electric terms. In this limit, electric field configurations become static and, therefore, the sum over the time direction gives an overall factor $\beta$. We thus get the purely 2D model 
\begin{equation}
S_{\rm ind}= \sum_x {1\over 2 e_{\rm eff}^2} \left( \ell^2 F_i + 2\pi M_i \right)^2 \ ,
\label{ind2}
\end {equation}
where
\begin{equation}
e_{\rm eff}^2 = {2\pi^2 \over \mu \beta v} {1\over \eta g} \ .
\label{effcou}
\end{equation}
This is the extreme anisotropic limit of Polyakov's compact QED action \cite{polyakov, polyakovbook}. In absence of the integer topological defects $M_i$ this would be simply the usual Maxwell action for electric fields only. In the vortex condensation phase, however, the integer topological defects become dynamical variables over which one has to sum in the partition function. 

It is important here to stress an often overlooked point. Electric charges are Noether charges and, as such, are locally conserved. Vortices, by contrast, are topological excitations: their conservation law is only global, and it can be locally violated by instanton events. In the 3D XY model, typically invoked in the phase-only treatment of the SIT, see e.g. \cite{fisher2}, the vortex current is locally conserved and a vortex condensate amounts only to the proliferation of large vortex loops. However, when vortices are coupled to an emergent gauge field, as required by a local formulation of the topological mutual statistics interactions, and the usual charge gauge symmetry is compact, local conservation or its absence becomes a dynamical question because instantons are present \cite{dst2}. When vortices have a finite tension, as in the Bose metal phase, instanton anti-instanton pairs are bound into tightly confined dipoles and are effectively suppressed. In this regime the vortex current is locally conserved, just like the electric current. In the condensation phase, however, the vortex lines connecting instanton anti-instanton pairs become tensionless, and thus unobservable Dirac strings, allowing instantons to proliferate freely \cite{dst2}. The vortex condensation phase is therefore not only a phase of long, fluctuating vortex loops, but primarily an instanton plasma. As a consequence, in this phase, the sum over magnetic topological defects $M_i$ runs over all unconstrained integers. We will now explore the consequences. 

The magnetic topological defects implement the periodicity under shifts of electric fields by integer multiples of $2\pi$, which, in turn, encodes a compact gauge group U(1) rather than the usual non-compact ${\mathbb R}$ that characterizes the electromagnetic response of normal insulators. The integration over the gauge fields in the partition function must be correspondingly restricted to the fundamental domain $[ -\pi/\ell, +\pi/\ell ]$. For the sake of simplicity, in the rest of this section we will set $\ell=1$, when not otherwise stated. Also, we keep the product $\beta v$ constant, which implies the previously discussed divergence of $\varepsilon$ near the SIT at zero temperature. We choose this constant as $8\pi^3 /\mu$ so that 
\begin{equation}
e_{\rm eff}^2 = {1\over 4\pi \eta(\beta) g} \ .
\label{corre}
\end{equation}

\begin{figure}[t!]
	\includegraphics[width=12cm]{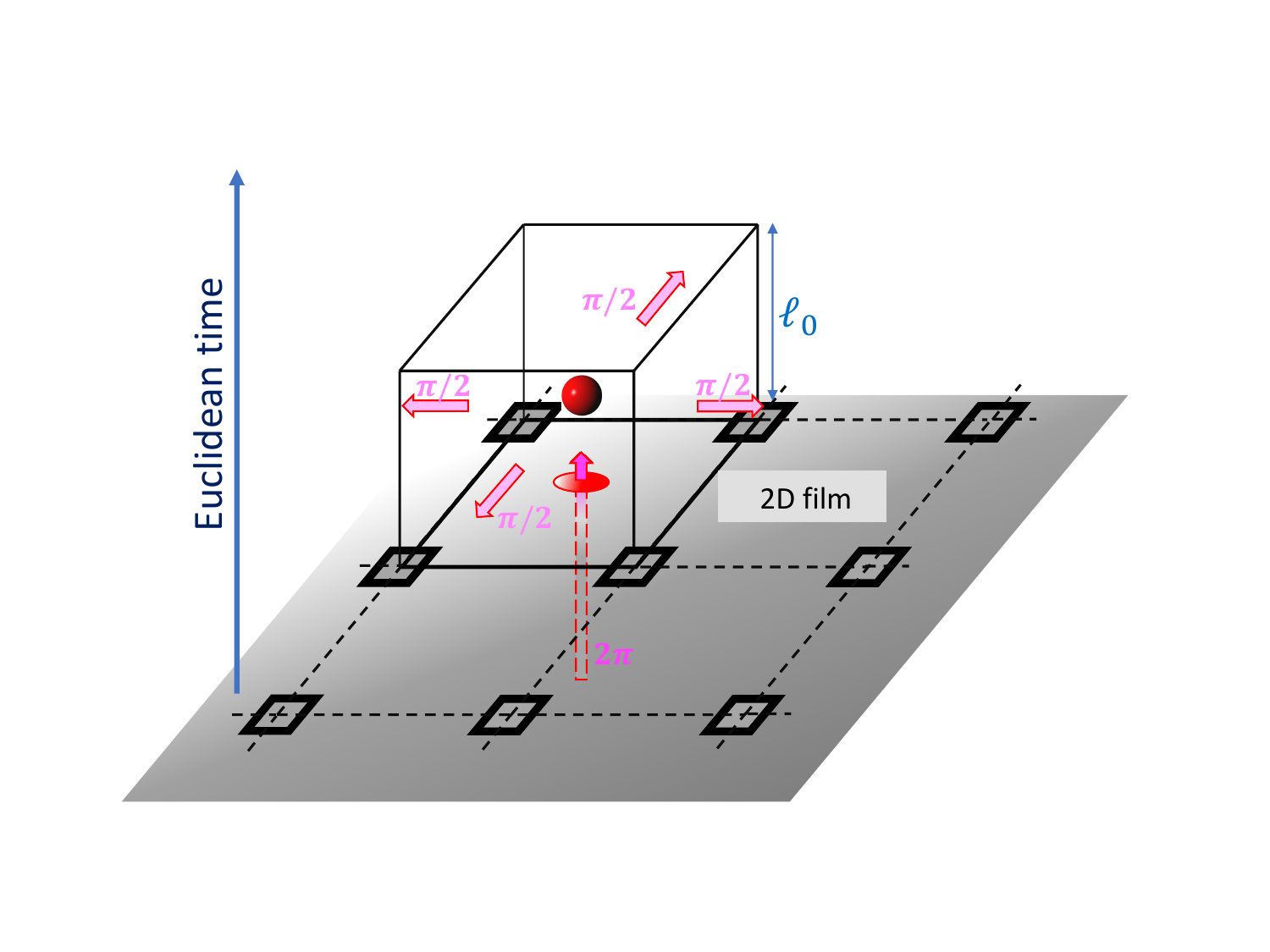}
	%\includegraphics[width=\textwidth]{phase_diagram.pdf}
	%\vspace{-1cm}
	\caption{A magnetic monopole representing a quantum tunnelling event in which a fundamental vortex of flux $2\pi$ disappears, radiating out its flux in the plane over and interval $\ell_0 = \ell/v$.}
	\label{fig:Fig.4}
\end{figure}

To show what is the consequence of the compact group, we decompose the topological defects into their longitudinal and transverse parts \cite{polyakov, polyakovbook}
\begin{equation}
M_i = \hat k_{i0} \left( N_0+ \xi_0 \right) + \hat k_{ij}  \left( N_j+ \xi_j \right) + d_i \chi \ ,
\label{decomp}
\end{equation}
where $N_{0,i}$ are integers, while $\xi_{0,i}$ and $\chi$ are real. The transverse and longitudinal components of the integers $M_i$ need not be separately integers, they must only sum to integers. However, when applying a finite difference operator on $M_i$, the transverse component falls out and, since differences of integers are still integers, the longitudinal components are restricted by the equation
\begin{equation}
d_i M_i = \nabla^2 \chi = m 
\label{mon1}
\end{equation}
with $m \in {\mathbb Z}$. 
Summing over the integers $N_{0,i}$ in the partition function shifts the integration domain of the original gauge fields back from a circle to the entire real line. Then, the real $\xi_{0,i}$ can be reabsorbed into the gauge fields, which are now real variables anyway. This gives back the action $S_{\rm nc}$ of the usual non-compact electromagnetism and, thus, only the longitudinal components of $M_i$ encode the physical topological excitations. Since $F_i$ is purely transverse and the remaining $d_i \chi$ is purely longitudinal, the cross-product term in the action vanishes and we get
\begin{eqnarray} 
S_{\rm ind} &&= S_{\rm nc} + S_{\rm top} \ ,
\nonumber \\
S_{\rm top} &&= 2\pi \eta(\beta) g \sum_{\{{\bf x}\}} m {1\over -\nabla^2} m  \ ,.
\label{topac}
\end{eqnarray}
The additional action due to the gauge group compactness (in the regime $v\ll1$) corresponds formally to the 2D statistical mechanics model of a Coulomb gas (for a review see \cite{minnhagen}) of point-like topological defects $m$, like in the XY model. Contrary to the XY model, however these are not vortices. Let us first recall that the dual gauge fields $F_{\mu}$ in 3D Euclidean spacetime play the same role as the usual magnetic fields $B_{\mu}$ in (3+1)-dimensional Minkowski space-time. Therefore, from this point of view, $m= d_i M_i$ can be viewed as magnetic monopoles emanating their magnetic fields on the 2D plane. Since $m=-d_0 M_0$, this can be viewed in (2+1)-dimensional Minkowski spacetime as a tunnelling event (instanton) in which a vortex appears or disappears from the vacuum over a time interval $\ell_0= \ell / v$, as shown in Fig. \ref{fig:Fig.4}.  In a condensate, however, vortices become tensionless and lose their status as physical, observable excitations, exactly as single charges lose their meaning in a superconducting condensate. Here, the original vortices survive only as unobservable Dirac strings.

This leads to an alternative, quite suggestive interpretation of these topological excitations in (2+1)-dimensional Minkowski space. Changing magnetic flux is always accompanied by circular electric fields around it. When the tunnelling time diverges because of a very small $v$, these become static electric loops. That this is indeed so is confirmed by the fact that $F_i = \epsilon^{ij} E^j$ Therefore $M_i$ correspond to closed, integer electric loops $E^i = \epsilon^{ij}M_j$ around the monopoles. These are formed and vanish again over the tunnelling time but become static electric vortices in the $v\to 0$ limit. These electric vortices are the dual of the magnetic vortices of the XY model. 

A 2D Coulomb gas undergoes a Berezinskii-Kosterlitz-Thouless (BKT) phase transition \cite{ber, kt1, kt2} as a function of its coupling constant. In the present case, the magnetic monopoles/ electric loops are free excitations for $\eta(\beta) g <1$ and logarithmically confined otherwise (for a review see \cite{minnhagen}). At finite temperatures, governed by the factor $\eta (\beta)$, this is the usual thermal BKT transition, albeit in terms of electric, rather than magnetic vortices. At zero temperature this is a quantum BKT transition driven by the coupling $g$ \cite{qbkt}. Given the extreme anisotropy this is a quantum transition with a diverging dynamical exponent $z$ and can thus be easily mistaken for a disorder-driven quantum Griffiths transition, although it has nothing to do with disorder \cite{qbkt}. 
In the following we shall derive what is the character of the low $\eta(\beta)g$ phase with free monopoles. 

To do so we consider two external probe charges of strength $\pm q_{\rm ext}$ and compute how their interaction potential is changed by the monopoles. We consider the expectation value of the Wilson loop operator $W(C)$, where $C$ is a closed loop in 3D Euclidean space-time, 
\begin{widetext}
\begin{equation}
\langle W(C) \rangle = {1\over Z_{A_{\mu}, M_i}} \sum_{\{ M_i \} }  \int_{-\pi}^{+\pi}  {\cal D} A_{\mu} 
\ {\rm e}^{-{ {\eta(\beta) g\over 2\pi}  }  \ \sum_{\bf x} \left( F_i-2\pi M_i \right)^2 } {\rm e}^{iq_{\rm ext} \sum_C A_{\mu} } \ .
\label{wilson1}
\end{equation}
\end{widetext}
When the loop $C$ is restricted to a plane formed by the Euclidean time and one of the space coordinates, $\langle W(C)\rangle$ measures the potential between the two external probe charges $\pm q_{\rm ext}$.  A perimeter law indicates a short-range potential, while an area-law is tantamount to a linear interaction between the probe charges (see, e.g., \cite{polyakovbook}).  In our extreme anisotropic regime, the area is made essentially by a single plaquette in the Euclidean time direction. 

For large enough couplings, the action is peaked near $F_i=M_i$, allowing to compute the Wilson loop in the saddle-point approximation. Using the lattice Stokes' theorem, one rewrites Eq.\,(\ref{wilson1}) as
\begin{widetext}
\begin{equation}
\langle W(C) \rangle = {1\over Z_{A_{\mu}, M_i}} \sum_{\{ M_i \} }  \int_{-\pi}^{+\pi}  {\cal D} A_{\mu} 
\ {\rm e}^{-{\eta(\beta) g\over 2\pi }  \ \sum_{\bf x} \left( F_i-2\pi M_i \right)^2 } {\rm e}^{iq_{\rm ext} \sum_S S_i (F_i -2\pi M_i) } \ ,
\label{wilson2}
\end{equation}
\end{widetext}
where $S_i$ are unit vectors perpendicular to the plaquettes forming the surface $S$ encircled by the loop $C$ and vanish on all other plaquettes. We have also multiplied the Wilson loop operator by 1 in the form ${\rm exp} (-i 2\pi q_{\rm ext} M_i)$ on all plaquettes forming $S$. We now repeat the steps leading from (\ref{ind2}) to (\ref{topac}) and obtain 
\begin{equation}
\langle W(C) \rangle_{\rm top} = {1\over Z_{\rs m}} \sum_{\{ m \} }  \ e^{-{2\pi \eta(\beta) g} \sum_{\bf x} m_{\bf x} 
	{1\over -\nabla^2} m_{\bf x} } \ e^{ i2\pi q_{\rm ext} \sum_S \hat d_i S_i  {1\over -\nabla^2} m_{\bf x} } \ .
\label{wilson3}
\end{equation} 

To sum over the monopole plasma we follow Polyakov \cite{polyakov, polyakovbook} by introducing a scalar field $\chi $ and rewriting (\ref{wilson3}) as
\begin{equation} 
\langle W(C) \rangle_{\rm top} = {1\over Z_{\rs m, \chi}} \int {\cal D} \chi \ {\rm e}^{-{1\over 8\pi \eta(\beta) g}\sum_{\bf x} d_i \chi d_i \chi}
\sum_{N} {z^N\over N!} \sum_{x_1,\dots , x_N}  \sum_{m_1,\dots , m_n = \pm1} e^{i \sum_{\bf x} m ( \chi + q_{\rm ext} \eta)} \ ,
\label{chi}
\end{equation}
where the angle $\eta = 2\pi \hat d_i S_i/(-\nabla^2)$ represents a dipole sheet on the Wilson surface $S$ and the monopole fugacity $z$ is determined by the self-interaction as
\begin{equation}
z = e^{-{2\pi \eta(\beta) g } G(0) } \ ,
\label{fugacity}
\end{equation}
with $G(0)$ the regularized inverse of the 2D Laplacian at coinciding arguments. In (\ref{chi}) we also adopted the dilute monopole approximation, valid at sufficiently low $\eta (\beta) g$, in which one takes into account only single monopoles $m=\pm 1$. The sum can now be explicitly performed, with the result,
\begin{equation}
\langle W(C) \rangle_{\rm top} = {1\over Z_{\rs \chi}} \int {\cal D} \chi \ e^{-{1\over 8\pi \eta(\beta) g}\sum_{\bf x} d_i \chi d_i \chi
	+ {16\pi \eta(\beta) g} z \ (1- {\rm cos}(\chi + q_{\rm ext}\eta))} \ ,
\label{sinegordon1}
\end{equation}
By shifting the field $\chi$ by $-q_{\rm ext}$ and introducing $\mu^2 =  8\pi \eta(\beta) g z$ we can rewrite this as
\begin{widetext}
\begin{equation}
\langle W(C) \rangle = {1\over Z_{\rs \chi}} \int {\cal D} \chi \ e^{-{1\over 4\pi \eta(\beta) g}\sum_{\bf x} {1\over 2} d_i (\chi -q_{\rm ext} \eta )d_i (\chi - q_{\rm ext}\eta ) + \mu^2 (1- {\rm cos}(\chi ))} \ .
\label{sinegordon2}
\end{equation}
\end{widetext}
For sufficiently small $\eta(\beta) g$ this integral is dominated by the classical solution to the equation of motion
\begin{equation}
\nabla^2 \chi_{\rm cl} = q_{\rm ext} \nabla^2 \eta + \mu^2 {\rm sin} \ \chi_{\rm cl} \ .
\label{eqmo1}
\end{equation}
Far from the boundaries of the Wilson surface $S$, where $S_i \ne 0$, this reduces to a one-dimensional equation. Taking the Wilson surface $S$ to lie, e.g., in the $(t, x_2)$ plane, this gives,
\begin{equation}
\hat d_{x1} d_{x1} \chi_{\rm cl} = - 2\pi q_{\rm ext} \hat d_{x1} S_1 + \mu^2 {\rm sin} \chi_{\rm cl} \ .
\label{eqmo2}
\end{equation}
Following\,\cite{polyakov, polyakovbook} we solve this equation, for simplicity, in the continuum limit. In doing so, however, we must replace the dimensionless $\mu$ with $mv \mu$, where $mv$ is the inverse length cutoff in the theory, 
\begin{equation}
\partial_{x1} \partial_{x1} \chi_{\rm cl} = 2\pi q_{\rm ext} \delta^{\prime} (x_1) +(mv)^2 \mu^2 {\rm sin} \chi_{\rm cl} \ .
\label{eqmo3}
\end{equation}
For $q_{\rm ext} = 1$ (corresponding to Cooper pairs in our case), the classical solution with the boundary conditions $\chi_{\rm cl} \to 0$ for $|x_1| \to \infty$ is 
\begin{equation}
\chi_{\rm cl} = {\rm sign} (x_1) \ 4 \ {\rm arctan} \  {\rm e}^{-mv\mu |x_1|} \ .
\label{cooperpairs}
\end{equation}
Inserting this back in (\ref{sinegordon2}) we get formula 
\begin{equation}
\langle W(C) \rangle = {\rm e}^{-\sigma A} \ ,
\label{areal}
\end{equation}
where $A$ is the area of the surface $S$, in our case $A=\ell_0 L$ where $L$ is the length in the spatial direction. We thus obtain the potential between the two charges as
\begin{eqnarray}
V(L) &&= \sigma L \ ,
\nonumber \\
\sigma &&= {mv^2\over \ell} \sqrt{8 z\over \pi \eta(\beta) g} \ ,
\label{strten}
\end{eqnarray}
where we have reinstated the lattice spacing $\ell$. This linear potential represents a string of electric field between the two charges, with tension $\sigma$ (energy per unit length) expressed in terms of the energy cutoff $mv^2$. This string prevents the two charges to be separated at will: at a certain distance it will be energetically favourable to pop out of the vacuum a Cooper pair and a Cooper hole and form two shorter strings. This is the essence of confinement and is the fundamental reason why the resistance is infinite for sufficiently large systems, even at finite temperatures up to the charge BKT transition. This {\it superinsulating state} is the exact dual of a superconductor, in which the conductance is infinite for finite temperatures up to the usual vortex BKT transition. 

Strictly speaking, however, this is true only in large enough systems because the sine-Gordon model is asymptotically free (for a review \cite{mussardo}). 
Confinement is due to the cosine interaction in the sine-Gordon model (\ref{sinegordon2}) becoming a relevant operator in the infrared. The monopole fugacity, however, is a running coupling that decreases at small scales. There is a dynamically generated scale $1/\sqrt{\sigma}$ below which the monopole-induced cosine interaction becomes effectively irrelevant under the renormalization group and the sine-Gordon theory crosses over to its asymptotically free Gaussian regime. Consequently, the effective string tension is negligible below that scale. Essentially the string becomes ``slack" below the scale $1/\sqrt{\sigma}$ and charges become, for all purposes, near free if the range of the Coulomb interaction is even smaller. This leads to the prediction of the existence of a scale below which the resistance of superinsulators should show a metallic saturation to a constant. 

The expression (\ref{strten}) for the string tension is valid in the interior of the superinsulating phase. The tension vanishes when approaching the BKT transition; the confinement of the monopoles/electric vortices corresponds to the liberation of electric charges. However, the behaviour of the tension near the transition is subtle and will be discussed below. Note also that, when approaching the zero-temperature quantum transition at $v\to 0$, $mv$ remains finite but the length scale $\ell$ vanishes so that $\ell/v$ remains finite, i.e. we reach a continuum 2D Euclidean field theory with a finite energy scale. This is the dimensional reduction implied by a diverging dielectric constant \cite{qbkt}. Finally let us stress that not only Cooper pairs are confined, but also single-charge quasiparticles, i.e. electrons \cite{poccia}.

\section{Josephson junction arrays} 

Josephson junction arrays (JJA) are square arrays of unit spacing of superconducting islands with nearest neighbour Josephson coupling $E_J$, ground capacitances $C_0$ and nearest neighbours capacitances $C$ \cite{fazio} (for a review see \cite{jjarev}). Their Hamiltonian is given by 
\begin{equation}
H =\sum_{\bf x} \ {C_0\over 2} V_{\bf x}^2 + \sum _{<{\bf x \bf y}>}
{C\over 2} \left( V_{\bf y}-V_{\bf x} \right) ^2 
+ E_{\rs J}\left( 1-{\rm cos}\ \left( \varphi _{\bf y} - \varphi _{\bf x} \right) \right) \ ,
\label{hamvolt}
\end{equation}
where boldface characters denote the sites of the two-dimensional array, $<{\bf x \bf y}>$ indicates nearest neighbours, $V_{\bf x}$ is the electric potential of the island at ${\bf x}$ and $\varphi _{\bf x}$ the phase of its order parameter. This Hamiltonian can be rewritten as
\begin{equation}
H= \sum_{\bf x} \ {1\over 2} V_{\bf x} \left( C_0 - C\nabla^2 \right) V_{\bf x} +
\sum _{{\bf x},i} \ E_{\rs J} \left( 1-{\rm cos} \ \left( d_i \varphi_{\bf x} \right)
\right) \ ,
\label{hamchar}
\end{equation}
The phases $\varphi _{\bf x}$ are quantum-mechanically conjugated to the charges on the islands: these are quantized in integer multiples of 2e (Cooper pairs),
$2e q_{\bf x}\ ,\, q_{\bf x} \in Z $, where $e$ is the electron charge. The Hamiltonian (\ref{hamchar}) can be written in terms of charges and phases by expressing the electric potentials $V_{\bf x}$ via a discrete version of Poisson's equation,
\begin{equation}
\left( C_0 -C\nabla^2 \right) V_{\bf x} = 2eq_{\bf x} \ .
\label{poisson}
\end{equation}
Using this in (\ref{hamchar}) we get
\begin{equation}
H= \sum_{\bf x} \ 4 E_{\rs C}\ q_{\bf x} {1\over {C_0/C}-\nabla^2 } q_{\bf x} +
\sum _{{\bf x},i} \ E_{\rs J} \left( 1-{\rm cos} \ \left( d _i \varphi_{\bf x} \right)
\right) \ ,
\label{hamfin}
\end{equation} 
where $E_{\rm C}\equiv e^2/2C$. The integer charges $q_{\bf x}$ interact via a two-dimensional Yukawa potential of ``mass" $\sqrt{C_0/C}$. The inverse of this mass plays the same role as the electric screening role $\lambda_s$ that we encountered when discussing the Keldysh potential in superconducting films. When this quantity is larger than the array size, the charge interaction becomes here too a two-dimensional Coulomb law. From now on we shall consider this nearest-neighbours capacitance limit $C\gg C_0$. In this regime, the charging energy $E_{\rm C}$ and the Josephson coupling $E_{\rm J}$ are the two relevant energy scales for the array. They can be traded for one energy $\omega_{\rm P}$, the array plasma frequency and one dimensionless parameter $g$ the dimensionless conductance,
\begin{eqnarray}
\omega_{\rm P} &&=\sqrt{8E_{\rs C}E_{\rs J}} \ ,
\nonumber \\
g &&=\sqrt{\pi ^2 E_{\rm J}\over 2E_{\rm C}} \ .
\label{JJApar}
\end{eqnarray}
In the following we will consider the physics of JJA at energies much below the plasma frequency, which has the role of the natural ultraviolet (UV) cutoff in the theory. 

The partition function of the JJA admits a phase-space path-integral representation involving an action. In this action time has to be considered also as discrete, as generally appropriate when degrees of freedom can change only in integer steps. We introduce thus a discrete time step $\ell_0$ whose inverse represents the energy scale of typical processes involving changes in charge degrees of freedom. We thus replace the time integrals and space sums over a lattice with nodes ${\bf x}$ by a sum over space-time lattice nodes $x$, with $x^0=t$ denoting the discrete time direction. Denoting by $d_0$ the (forward) finite time differences we obtain
\begin{eqnarray}
Z &&=\sum_{\{q\}} \int_{-\pi }^{+\pi } {\cal D}\varphi
\ {\rm exp}(-S)\ ,
\nonumber  \\
S&&=\sum _x  i\ q_x d_0 \varphi_x + 4 \ell_0 E_{\rs C} \ q_x {1\over -\nabla^2 } q_x% \nonumber \\
+\sum _{x, i} \ell_0 E_{\rs J} \left( 1- {\rm cos}\ \left( d_i \varphi_x \right) \right) \ ,
\label{partjja}
\end{eqnarray} 

We now use the Villain representation (for a review see \cite{kleinertbook}) to express the cosine interaction in (\ref{partjja}) in terms of a set of integer link variables $a_i$. We also introduce the notation $j_0$ for the integer charges and $j_i$ for real charge currents, 
\begin{widetext}
\begin{eqnarray}
Z&&=\sum_{\{ a_i \}, \{ j_0 \} }   \int {\cal D} j_i \int_{-\pi }^{+\pi } {\cal D}\varphi
\ {\rm exp}(-S)\ ,
\nonumber \\
S&&=\sum _x  i\ j_0 \Delta_0 \varphi + i j_i \left( \Delta_i \varphi + 2\pi a_i \right) 
+ 4 \ell_0 E_{\rs C} \ j_0{1\over -\Delta } j_0 + {1\over 2\ell_0 E_{\rs J}} {j_i}^2 \ ,
\label{partition1}
\end{eqnarray} 
\end{widetext}
where we have dropped the underscripts referring to the lattice positions of the variables.

As was first pointed out in \cite{dst}, this JJA model still misses a crucial piece, a kinetic term for the vortices, encoded in the integer variable $a_i$ conjugate to the charge current. We have already pointed out previously how the vortices in this model are coreless XY vortices, which can move freely without dissipation. This is confirmed by the simple fact that integrating over charge fluctuations leads automatically to a vortex kinetic term \cite{fazio}. It is a general and fundamental principle of field theory that whatever is induced radiatively by fluctuations {\it must} be included at bare level. The vortex kinetic term cannot be left out from (\ref{partition1}). 
Such a kinetic term would involve the time derivative of $a_i$ and would represent phase slips corresponding to one vortex moving from one plaquette of the lattice to a neighbouring one. When the coefficient of this kinetic term $(\partial_0 a_i)^2$ takes the value $\pi^2 /4\ell_0 E_C$ we can introduce a real Lagrange multiplier $a_0$ and a fictitious ``electric field" $\phi_i = k_{i0} a_{0}+k_{ij}a_j$ and write the vortex kinetic term and the charge Coulomb interaction compactly as
\begin{widetext}
\begin{eqnarray}
Z &&=\sum_{\{ a_i \}, \{ j_0 \} }   \int {\cal D} a_0  {\cal D} j_i  \int_{-\pi }^{+\pi } {\cal D}\varphi
\ {\rm exp}(-S)\ ,
\nonumber \\
S &&=\sum _x  i\ j_0 \left( d_0 \varphi +2\pi a_0\right)  + i j_i \left( d_i \varphi + 2\pi a_i \right) 
+ {1\over 2\ell_0 E_{\rs J}} {j_i}^2 + {\pi^2 \over 4\ell_0 E_C} \phi_i^2 \ .
\label{partition2}
\end{eqnarray} 
\end{widetext}
In this representation, the Coulomb interaction between the charges follows from the Gauss law constraint associated with the Lagrange multiplier $a_0$. This procedure works only at a particular value of the vortex kinetic term that makes the model self-dual  under the interchange of charge and vortex variables, while simultaneously interchanging $\pi^2 E_J\leftrightarrow 2E_C$, or alternatively, $g \leftrightarrow 1/g$. As a consequence it was called the self-dual approximation in\,\cite{dst}. We shall adopt this approximation here too. 

Since the charge current $j_{\mu}$ is conserved it can be represented as the (dual) field strength associated to a second fictitious gauge field $b_{\mu}$ as $j_0 = k_{0i}b_i$, $j_i= k_{i0}b_0 + k_{ij}b_j$, where $b_0$ is a real variable while $b_i$ are integers. We then use Poisson's formula,
\begin{equation}
\sum_{n_{\mu}} f\left( n_{\mu} \right) = \sum_{k_{\mu}} \int dn_{\mu} f\left( n_{\mu} \right) {\rm e}^{i2\pi n_{\mu}k_{\mu}} \ ,
\label{poisson}
\end{equation}
to turn a sum over integers $\{ n_{\mu} \}$ into an integral over real variables, making all components of the gauge fields $a_{\mu}$ and $b_{\mu}$ real at the price of introducing integer link variables $Q_i$ and $M_i$,
\begin{widetext}
\begin{eqnarray}
Z &&= \sum_{\{ Q_i \} }\sum_{\{ M_i \} }  \int {\cal D} a_{\mu} {\cal D} b_{\mu} \int_{-\pi }^{+\pi } {\cal D}\varphi \ {\rm exp}(-S)\ ,
\nonumber \\
S &&=\sum _{x}  i 2\pi \ a_{\mu} k_{\mu \nu} b_{\nu}+{1\over 2\ell_0 E_{\rs J}} {j_i}^2 + {\pi^2 \over 4\ell_0 E_C} \phi_i^2 + i 2\pi a_i Q_i + i 2\pi b_i M_i 
+ b_i \left( \hat k_{i0} \Delta_0 \varphi + \hat k_{ij} \Delta_j\varphi \right) 
+b_0 \hat k_{0i}\Delta_i \varphi \ .
\label{partition3}
\end{eqnarray} 
\end{widetext}

Finally, we note that the quantities $(1/2\pi) \hat k_{\mu \nu} \Delta_{\nu} \varphi $ are the circulations of the array phases around the plaquettes orthogonal to the direction $\mu$ in 3D Euclidean space-time and are thus quantized as $2\pi \ {\rm integers}$. We can thus absorb the quantities $\left( \hat k_{i0} \Delta_0 \varphi + \hat k_{ij} \Delta_j\varphi \right)$ in a redefinition of the integers $M_i$ and, additionally define $\hat k_{0i} \Delta_i \varphi = 2\pi M_0$. The original integral over the phases $\varphi$ can then be traded for a sum over the vortex numbers $M_0$, 
\begin{widetext}
\begin{eqnarray}
Z &&= \sum_{\{ Q_i \} }\sum_{\{ M_{\mu}\} }  \int {\cal D} a_{\mu} {\cal D} b_{\mu} \ {\rm exp}(-S)\ ,
\nonumber \\
S &&=\sum _{x}  i 2\pi \ a_{\mu} k_{\mu \nu} b_{\nu}+{1\over 2\ell_0 E_{\rs J}} {j_i}^2 + {\pi^2 \over 4\ell_0 E_C} \phi_i^2 + i 2\pi a_i Q_i + i 2\pi b_{\mu} M_{\mu} \ .
\nonumber \\
\label{partition3}
\end{eqnarray} 
\end{widetext}
After a rescaling of the gauge fields by a factor of $1/2\pi$, this is exactly the same model (\ref{totac}) that we obtained for granular superconducting films on topology and symmetry grounds alone, with the translations
\begin{eqnarray}
mv^2 && \rightarrow \quad \omega_{\rm P} \ ,
\nonumber \\
g &&\rightarrow \sqrt{\pi^2 E_{\rm J} \over 2 E_{\rm C}} \ .
\label{translation}
\end{eqnarray}
All the results we obtained previously for films carry over verbatim to JJA. 

\section{Three dimensions: gauging vortices}

Superinsulation is a phase of superconductors with emergent granularity of the condensate. Such systems are not confined to thin films, they exist also in bulk materials \cite{parra}. In thin films, superinsulation arises because the large dielectric constant renders the electric interaction logarithmic, like the vortex interaction, so that they can compete. In 3D it is the other way around, it is the magnetic interaction between monopoles that is the same as the electric interaction between charges so that they can again compete. 

\begin{figure}[t!]
	\includegraphics[width=12cm]{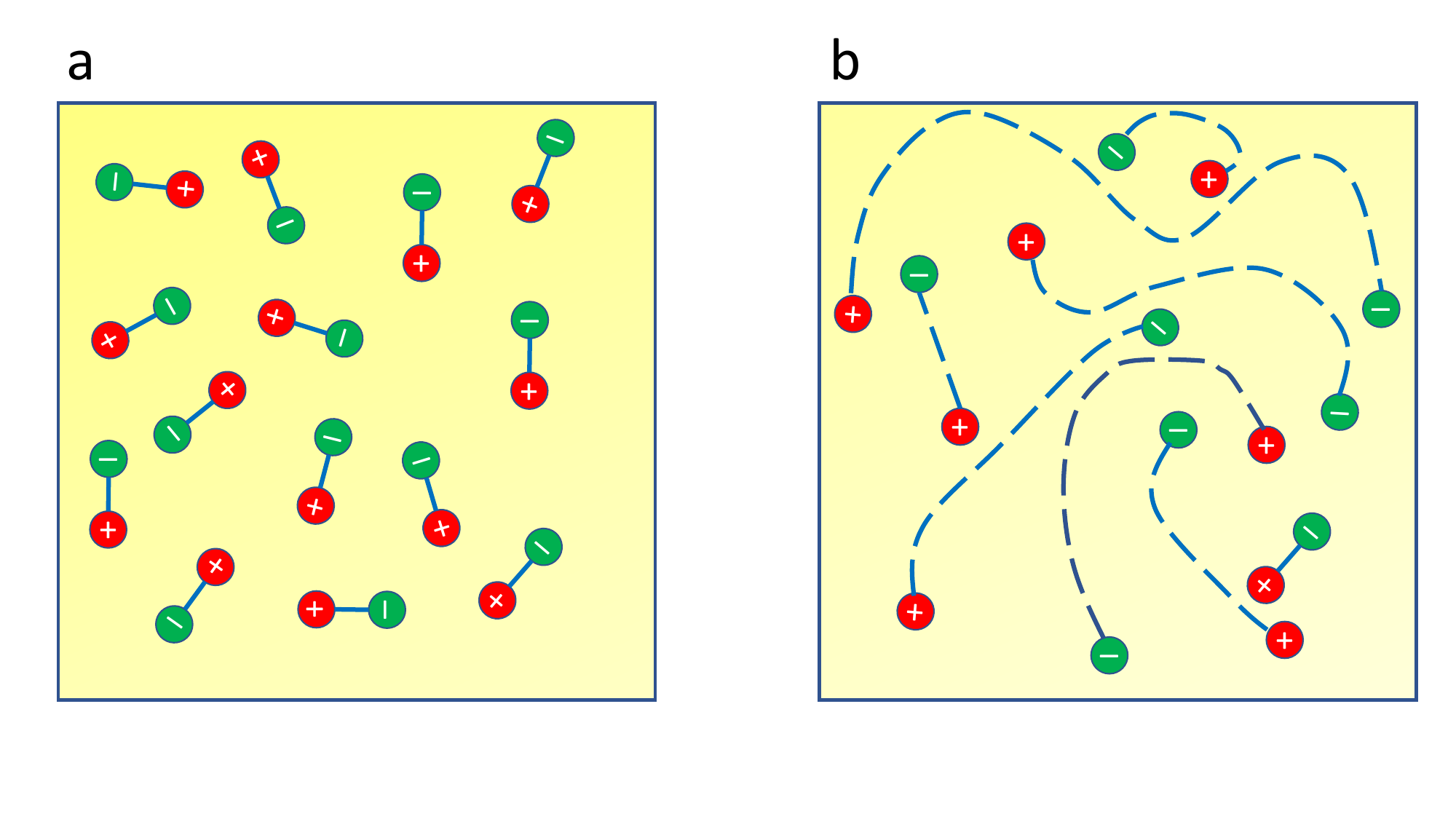}
	%\includegraphics[width=\textwidth]{phase_diagram.pdf}
	%\vspace{-1cm}
	\caption{(a) Short magnetic dipoles hold together by tensionful vortices. (b) Magnetic monopoles with their tensionless Dirac strings.}
	\label{fig:Fig.5}
\end{figure}

In 3D, the XY vortices between granules become 1D extended objects. We shall consider a model of charges (Cooper pairs) on the grains and open XY vortices in the spaces between them. The open vortices have couples of $\pm \pi/e$ magnetic monopoles at their ends. Actually, when the vortices have tension, they form only short magnetic dipoles; it is when the vortices become tensionless and condense that monopoles are liberated. When a vortex becomes tensionless we can move it and displace it at will with no energy price. For example we can take the vortex between two magnetic charges, make it infinitely long so that it runs from one magnetic charge to infinity and then comes back to the other magnetic charge. Since they are quantized, these two strands become simply two unobservable Dirac strings for two bona fide magnetic monopoles. Since these are bosons they condense, forming a superinsulating state, as shown in Fig. \ref{fig:Fig.5}.

In analogy to the 2D case, we formulate a model of emergent gauge fields coupled to charges and vortices on a 4D Euclidean lattice of spacing $\ell$. In this case, however, vortices are no more characterized by world-lines but, rather by world-sheets and couple, thus to antisymmetric tensor gauge fields. The lattice world sheets are described by antisymmetric integers $J_{\mu \nu}$ on the plaquettes forming the surface. We shall decompose these open world-sheets in their transverse bulk and their longitudinal boundary,
\begin{equation}
J_{\mu \nu} = M_{\mu \nu} +\left(  d_{\mu} C_{\nu} - d_{\nu} C_{\mu} \right) \ ,
\label{decompo}
\end{equation}
where $d_{\mu} M_{\mu \nu} = 0= d_{\nu} M_{\mu \nu}$. As in 2D, the presence of the monopoles world-lines $C_{\mu}$ at the end of vortices is a dynamical one which depends on the vortex tension. 
Since we are interested primarily in this bulk tension we will consider only the coupling of the transverse bulk components $M_{\mu \nu}$ to an antisymmetric gauge field $b_{\mu \nu}$. We will keep in mind that, when vortices become tensionless there is phase transition to a condensate of magnetic monopoles, in full analogy to the 2D case, with the only difference that here monopoles are particles and not instantons. 

The total action, including the coupling of the charges to the emergent gauge field $a_{\mu}$ with field strength $f_{\mu \nu}$, is given by 
\begin{equation}
	S = \sum_x {\ell ^4\over 4f^2}
	f_{\mu \nu} f_{\mu \nu} -i{\ell ^4\over 4\pi }
	a_{\mu }k_{\mu \alpha \beta }b_{\alpha \beta } + {\ell ^4\over 12 \Lambda^2}
	h_{\mu \nu \alpha }h_{\mu \nu \alpha} 
	+i\ell a_{\mu }Q_{\mu }
	+i\ell ^2{1  \over 2} b_{\mu \nu}M_{\mu \nu} \ ,
	\label{fullaction}
\end{equation}
where $k_{\mu \nu \alpha}$ is the lattice BF term, see Appendix A, $f$ is a dimensionless coupling and $\Lambda$ has the canonical dimension of 1/length. To describe materials with the relative electric permittivity $\varepsilon$ and relative magnetic permeability $\mu$, we incorporate the velocity of light $v=1/\sqrt{\varepsilon \mu} <1 $ by defining the Euclidean time lattice spacing as $\ell_0 = \ell/v$ and by rescaling all time derivatives, currents, and zero-components of gauge fields by the factor $1/v$, as in the previous section.  As a consequence, both gauge fields acquire a dispersion relation $E = \sqrt{m^2 v^4 + v^2 {\bf p}^2}$ with the topological mass \cite{bowick} $m= f\Lambda/2\pi v$. The dimensionless parameter $f=O(e)$ encodes the effective Coulomb interaction strength in the material, $\Lambda$ is the magnetic scale $\Lambda=O(1/\lambda_{\rs L})$, where $\lambda_{\rs L}$ is the London penetration depth of the superconducting grains. This action is the exact 3D analogue of the 2D action (\ref{totac}) derived above. 

As in the 2D case, we proceed to integrate out the fictitious gauge fields $a_{\mu}$ and $b_{\mu \nu}$ to obtain an induced action for the charges and vortices, 
\begin{equation}
	S_{\rm ind} = \sum_x {f^2 \over 2\ell^2} \ q_{\mu }{\delta _{\mu \nu }
		\over {(mv)^2- \Delta_4 }} q_{\nu } 
	+ {\Lambda^2 \over 8} \ m_{\mu \nu}
	{{\delta _{\mu \alpha } \delta _{\nu \beta }-\delta _{\mu \beta}
			\delta _{\nu \alpha }} \over {(mv)^2-\Delta_4}} m_{\alpha \beta } 
	+i {\pi (mv)^2 \over \ell}
	\ q_{\mu } {k_{\mu \alpha \beta}
		\over {\Delta_4 \left( (mv)^2- \Delta_4 \right) }} m_{\alpha \beta } \ ,
	\label{topeffac}
\end{equation}
The last term represents the Aharonov-Bohm phases of charged particles around 1D vortices of width $\lambda_{\rm L}$. On scales much larger than $\lambda_{\rm L}$, where the denominator reduces to $(mv)^2 \Delta_4$, this term becomes $i2\pi \ {\rm integer}$, as can be easily recognized by a computation analogous to the 2D case explained in detail above. Accordingly, we shall henceforth neglect this term. 

The other important consequence of the topological interaction, however, is that it induces a screening of interactions, in form of a mass for Cooper pairs and a tension for vortices. These are encoded in the short-range kernels in the action (\ref{topeffac}), which we shall approximate by constants. World-lines and world-surfaces 
are thus assigned ``energies" proportional to their length $N$ and area $A$ (measured in numbers of links and plaquettes),
\begin{eqnarray}
	S_{\rs N} = \pi (mv\ell) G \ {f\over \Lambda\ell} \ Q^2 N \ ,
	\nonumber \\
	S_{\rs A} = \pi (mv\ell) G \ {\Lambda\ell \over f} \  M^2 A \ ,
	\label{energies}
\end{eqnarray}
where $G=G(mv\ell)$, with $G(mv\ell)$ the diagonal element of the lattice kernel $G(x-y)$ representing the inverse of the operator $\ell^2 \left( (mv)^2 -\Delta_4 \right) $, and $Q$ and $M$ are the integer quantum numbers carried by the two kinds of topological defects. However, also the entropy of link strings and plaquette surfaces is proportional to their length and area \cite{nelson}, $ \mu_{\rm N} N$ and $\mu_{\rm A} A$. Both coefficients $\mu$ are non-universal: $\mu_{\rs N} \simeq {\rm ln}(7)$ since at each step the non-backtracking string can choose among 7 possible directions on how to continue, while $\mu_{\rs A}$ does not have such a simple interpretation but can be estimated numerically. This gives for both types of topological defects a ``free energy" proportional to their dimension and with coefficients that can be positive or negative depending on the parameters of the theory. The total free energy is 
\begin{equation}
	F  =\pi (mv\ell) G \left[ \left( {f\over \Lambda \ell} \ Q^2 -{1\over \eta_{\rm Q}} \right) N 
	+ \left( {\Lambda \ell \over f} \  M^2 -{1\over \eta_{\rm M}}\right) A \right] \ ,
	\label{freeenergy}
\end{equation} 
where we have defined 
\begin{equation}
	\eta_{\rm Q}=  {\pi (mv\ell) G\over \mu_{\rm N}} \ , \qquad \eta_{\rm M}=  {\pi (mv\ell) G \over \mu_{\rm A}} \ .
	\label{eta}
\end{equation}
If the coefficients are positive, the self-energy dominates; in this regime Cooper pairs and/or vortices are suppressed by their large action. 
If the coefficients, instead are negative, the entropy dominates and large configurations are favoured in the free energy.  The phase in which long world-lines of Cooper pairs dominate the Euclidean partition function is a charge Bose condensate, i.e. a global superconductor, in which phase coherence is established over all local superconducting grains in the material. For vortices, proliferation of large world-surfaces means that the small dipoles become  a condensate of free magnetic monopoles with unobservable, tensionless Dirac strings between them. 

The combined energy-entropy balance equations are again best viewed as defining the interior of an ellipse on a 2D integer lattice of electric and magnetic quantum numbers, 
\begin{equation}
	{Q^2 \over r_{\rm Q}^2} + {M^2 \over r_{\rm M}^2} < 1 \ ,
	\label{ellipse}
\end{equation}
where the semiaxes are given by
\begin{eqnarray}
	r_{\rm Q}^2  = {\ell \Lambda \over f} \ {1\over \eta_{\rm Q}} = {\ell \Lambda \over f} \sqrt{\mu_{\rm N}\over \mu_{\rm A}} \ {1\over \eta} \ ,
	\nonumber \\
	r_{\rm M}^2  = {f\over \ell \Lambda} \ {1\over \eta_{\rm M}} = {f\over \ell \Lambda} \sqrt{\mu_A\over \mu_{\rm N}} {1\over \eta} \ ,
	\label{semi}
\end{eqnarray}
with 
\begin{equation}
	\eta= \sqrt{\eta_{\rm Q} \eta_{\rm M}}=\pi (mv\ell) G/\sqrt{\mu_{\rm N} \mu_{\rm A}} \ . 
	\label{etasingle}
\end{equation}
The phase diagram is established by determining which integer charges lie within the ellipse when the semi-axes are varied. It can be expressed as a function of 
the parameters $\eta$ and $g= (\ell \Lambda/f) \sqrt{\mu_{\rm A}/\mu_{\rm N}}$. 
\begin{eqnarray}
	&&\eta < 1 \to 
	\begin{cases}
		g> 1 \ , {\rm superconductor} \ ,\\
		g< 1\ , {\rm superinsulator} \ ,\\
	\end{cases}
	\nonumber \\
	&&\eta> 1 \to
	\begin{cases}
		g > {\eta}
		\ , {\rm superconductor} \ ,\\
		{1\over \eta}
		<g<\eta \ , {\rm bosonic\ topological\ insulator} \ ,\\
		g < {1\over \eta} \ , {\rm superinsulator }\ .\\
	\end{cases}
	\label{phasestructure}
\end{eqnarray}

\begin{figure}[t!]
	\includegraphics[width=12cm]{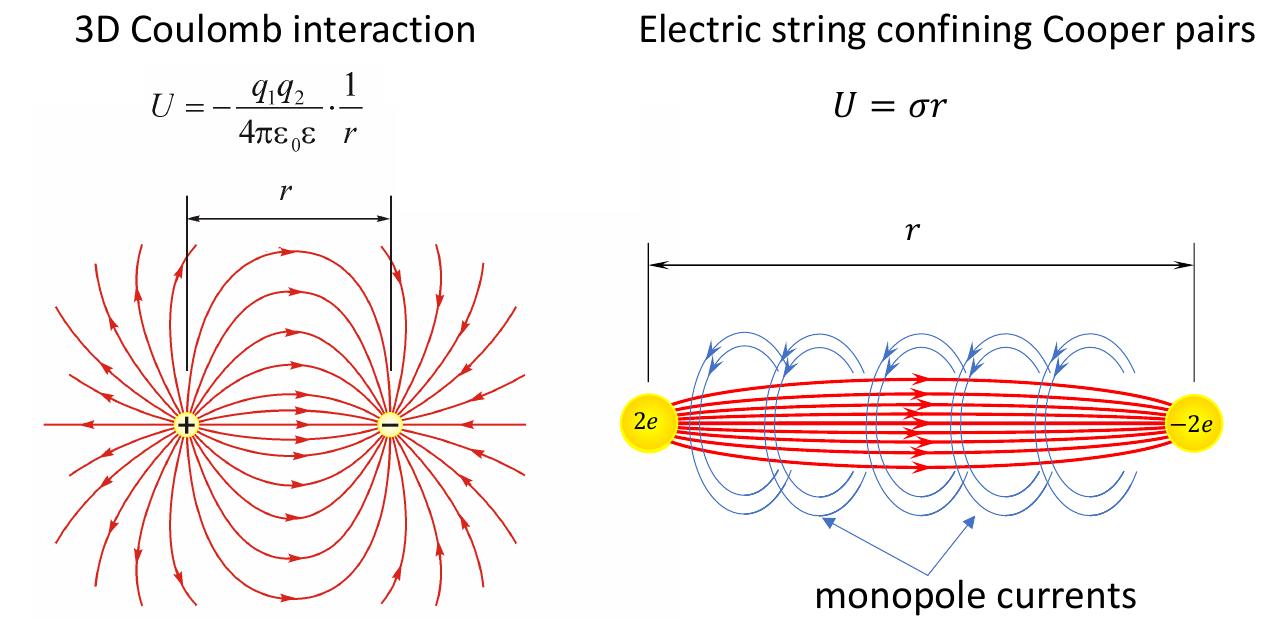}
	%\includegraphics[width=\textwidth]{phase_diagram.pdf}
	%\vspace{-1cm}
	\caption{The magnetic monopole condensate squeezes electric fields into flux tubes; these imply a linear potential with string tension $\sigma$.}
	\label{fig:Fig.6}
\end{figure}

For small $g$, i.e. for large values of the effective electric interaction $f$, a condensate of magnetic monopoles can form. With a Wilson loop calculation fully analogous to the 2D case we presented in detail above, one can infer that also in 3D the monopole condensate squeezes electric fields into flux tubes forming strings and the corresponding linear potential for probe charges inserted in the material, as shown in Fig. \ref{fig:Fig.6}. This is a bulk superinsulating state.

\section{Experiments}
\subsection{Resistance scaling} 
\begin{figure}[t!]
	\includegraphics[width=12cm]{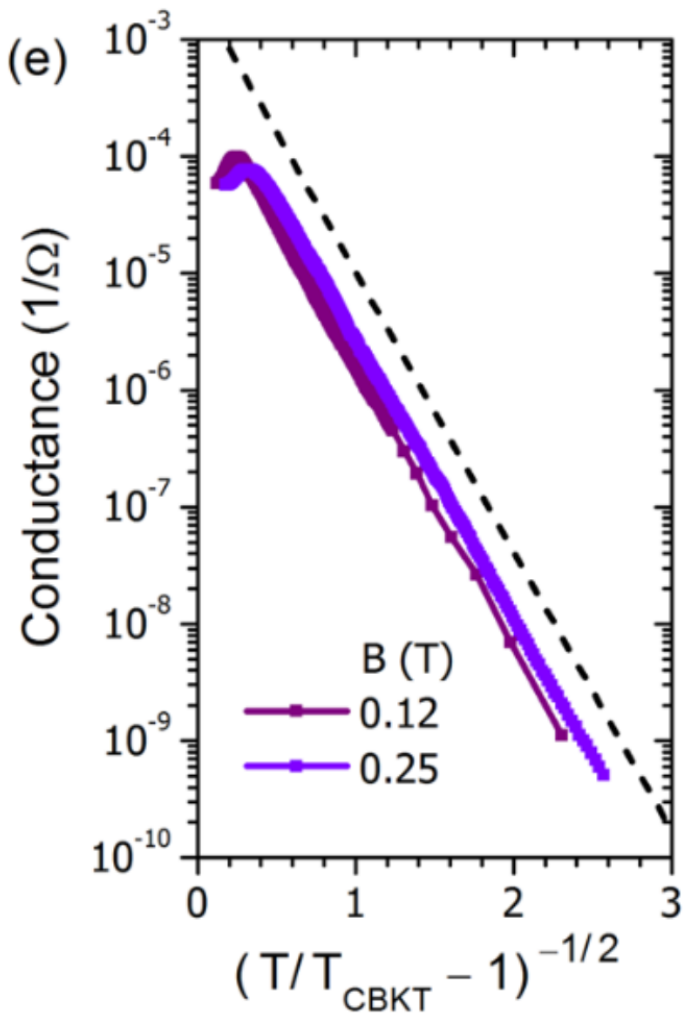}
	\vspace{-3cm}
	\caption{Conductance of a NbTiN film in logarithmic scale as a function of $(T/T_C-1)^{-1/2}$ near the superinsulating transition. A straight line like the dotted one indicates perfect BKT scaling. The experimental values for two values of applied magnetic field, used to drive the transition, are shown in two shades of violet. From ref. \cite{nbtin}. }
	\label{fig:Fig.7}
\end{figure}

The resistance scaling at the deconfinement transition is obtained from the behaviour of the string tension $\sigma $ when approaching the critical temperature \cite{vft}. The quantity $\xi = 1/\sqrt{\sigma} $ is a measure of the typical size of a linearly bound charge-hole pair. For $r < \xi$ the string is slack and charges are free,  while they start to feel the string around $r \approx \xi$. Thus, the number of free charge carriers available for transport and the conductance scale like $\xi^{-D}$. In 2D, $\xi$ exhibits BKT scaling while in 3D it shows an even more singular divergence at the transition \cite{vft}. This leads to the 2D BKT resistance scaling
\begin{equation}
R_{\rm 2D} \propto {\rm e}^{\sqrt{b_{\pm}\over |T-T_{\rm BKT}|}} \ ,
\label{2Dscal}
\end{equation}
where $b_{\pm}$ are constants that can differ below and above the transition. In 3D, instead one obtains the Vogel-Fulcher-Tamman (VFT) resistance scaling
\begin{equation}
R_{\rm 3D} \propto {\rm e}^{b_{\pm}\over |T-T_{\rm VFT}|} \ .
\label{2Dscal}
\end{equation}
The BKT scaling of the conductance near the deconfinement transition has been clearly observed in TiN \cite{vinokurnature} and NbTiN  \cite{nbtin} films, as shown in Fig. \ref{fig:Fig.7}. The 3D VFT scaling was observed in InO films \cite{shahar}, which are indeed thicker. 

\begin{figure}[t!]
	\includegraphics[width=12cm]{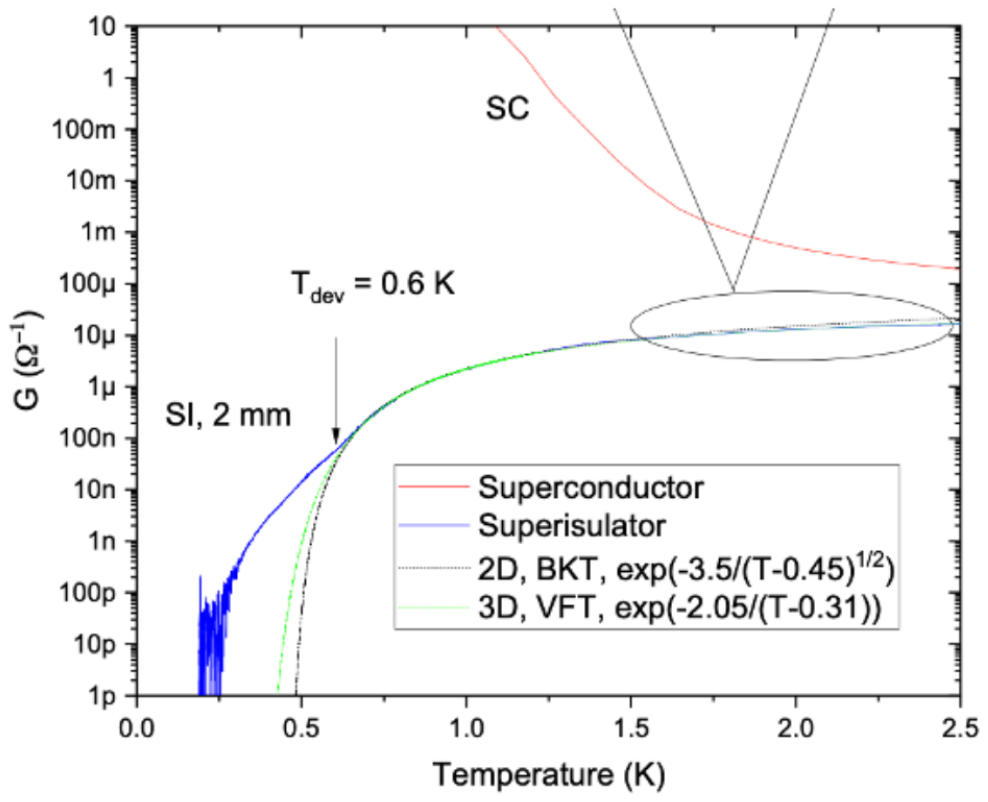}
	\vspace{-3cm}
	\caption{Conductance of a nanopatterned NbTiN film in logarithmic scale as a function of temperature.  In red the superconducting behaviour before nanopatterning. In blue the superinsulating behaviour after nanopatterning. In green the theoretical VFT scaling. The observed VFT scaling persist right until finite size effects are expected to distort the curve. This is an example of superinsulation in an ordered structure. From ref. \cite{bsi}.}
	\label{fig:Fig.8}
\end{figure}

VFT scaling of the conductance was observed also in a nanopatterned NbTin film \cite{bsi}, as shown in Fig. \ref{fig:Fig.8}. This structure was obtained by punching a regular array of holes in an otherwise superconducting film, essentially ``simulating" an extremely large JJA. The VFT scaling of the conductance persists right down to temperatures where finite-size effects are expected to distort the curve. This is a particularly important experiment since it shows superinsulation in a perfectly ordered structure, dispelling any notion of disorder as the root cause of the phenomenon. 

\subsection{Why superinsulation is harder to observe than superconductivity near the transition} 
Superinsulation has not yet been observed in synthetic Josephson junction arrays, and the expected BKT (or VFT) scaling is significantly harder to resolve in thin films than its superconducting counterpart. The reason is straightforward. In superconducting films, the relevant excitations, vortices, interact via a logarithmic potential. This leads to a correlation length that, while large near the transition, typically remains within experimentally accessible system sizes. As a result, the critical scaling can be observed over a broad temperature range. In contrast, in the superinsulating phase, charges are bound by a linear confining potential. This interaction is much stronger than the logarithmic binding of vortices and leads to a much more rapidly growing correlation length as the transition is approached. Consequently, the characteristic length scale governing transport quickly exceeds the lateral size of realistic samples. This has a direct experimental implication: finite-size effects become dominant much earlier in superinsulators than in superconductors. In practice, samples that are sufficiently large to display clear superconducting BKT scaling are often already too small to access the true critical regime of the dual superinsulating phase. As a result, the expected scaling behaviour is easily missed, not because it is absent, but because it occurs outside the accessible window. In particular, currently available synthetic Josephson junction arrays are too small to clearly exhibit superinsulating behaviour \cite{poccia}. However, when significantly larger structures are realized, for example through nanopatterning, the scaling signature of superinsulation emerges immediately.

\subsection{The electric Meissner effect and the IV curves}
\begin{figure}[t!]
	\includegraphics[width=12cm]{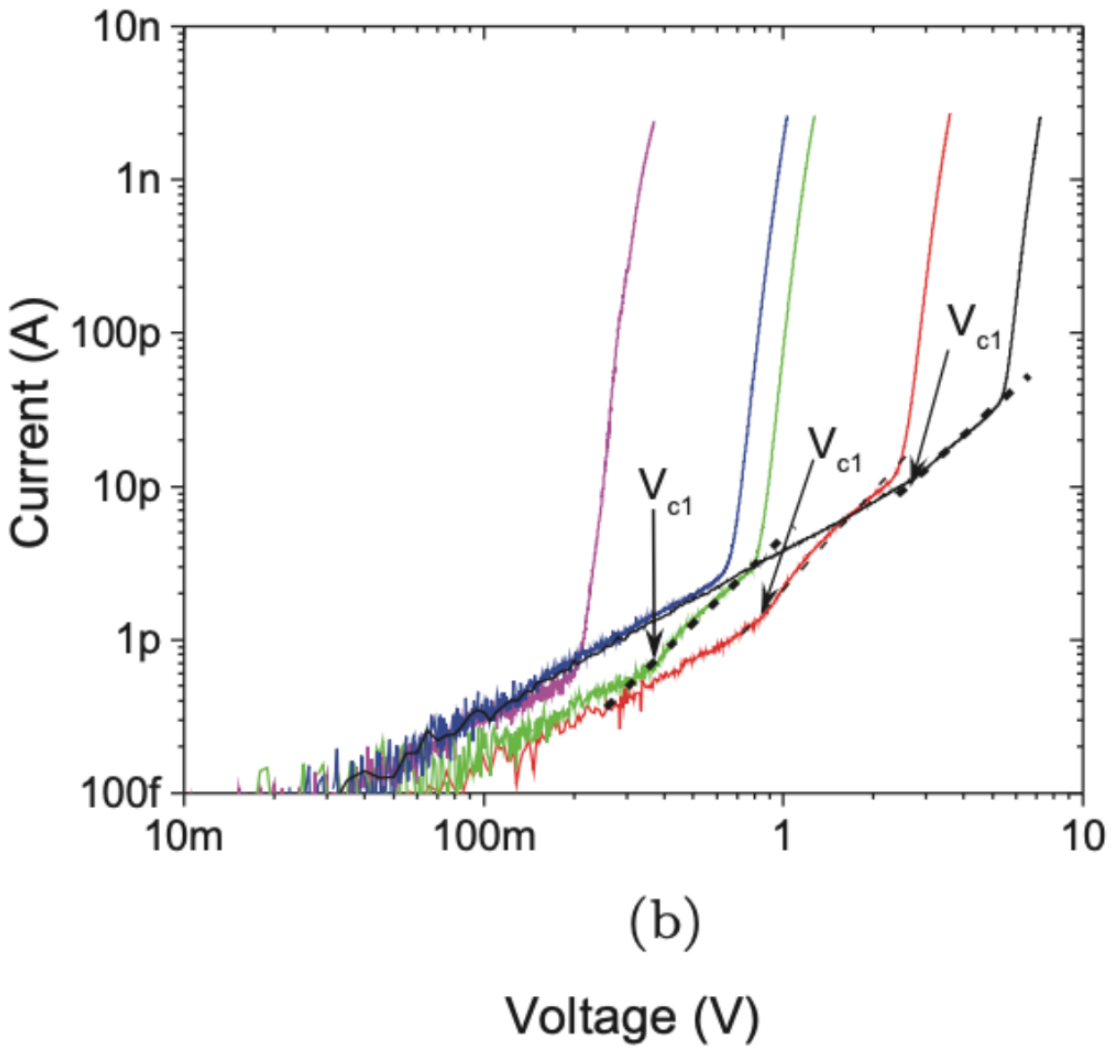}
	\vspace{-2cm}
	\caption{The I(V) curves of superinsulating NbTin films of different lateral sizes. Two samples (violet and blue curves) are too small to resolve the lower kink corresponding to the onset of the mixed state and show only the upper kink when superinsulation is destroyed. Both kinks, however, are clearly visible in the three larger samples. The critical voltage $V_{\rm c1}$ is the lower critical voltage denoting the onset of the mixed state. From \cite{electrostatics}.}
	\label{fig:Fig.9}
\end{figure}

When a constant electric field is applied to a superinsulating sample the effective string tension is reduced to $\sigma_{\rm E} = \sigma -2eE$. Correspondingly the string scale is increased to $r_{\rm E}= 1/\sqrt{\sigma_{\rm E}}$. Two cases can be realized in samples of lateral size $L$. If $r_{\rm E} < L$ the typical length of strings is smaller than the sample and the current is exponentially suppressed \cite{electrostatics}, 
\begin{equation}
I \propto V {\rm e}^{-{\sigma L\over k_{\rm B}T}} \ ,
\label{mei}
\end{equation}
where $V$ is the applied voltage. This is the electric Meissner state: electric fields are expelled and the current vanishes identically in infinite samples. If, instead $r_{\rm E} > L$, thin electric flux tubes can penetrate from one side of the sample to the other and current starts to flow with a power law \cite{electrostatics}
\begin{equation}
I \propto \left( V-{\sigma L\over 2e} \right)^{\tau (T)}  \ ,
\label{mix}
\end{equation} 
where $\tau (T)$ is a temperature-dependent exponent. This is the mixed state of superinsulators. Correspondingly one should expect two kinks in the I(V) characteristic curves, one at a lower critical voltage $V_{\rm c1}$ for the onset of the mixed state and one at higher critical voltage $V_{\rm c2}$ where superinsulation is destroyed. These two kinks are very well visible in the I(V) curves measured in the superinsulating state of a NbTin film, shown in Fig. \ref{fig:Fig.9}.

\subsection{Asymptotic freedom}

\begin{figure}[t!]
	\includegraphics[width=12cm]{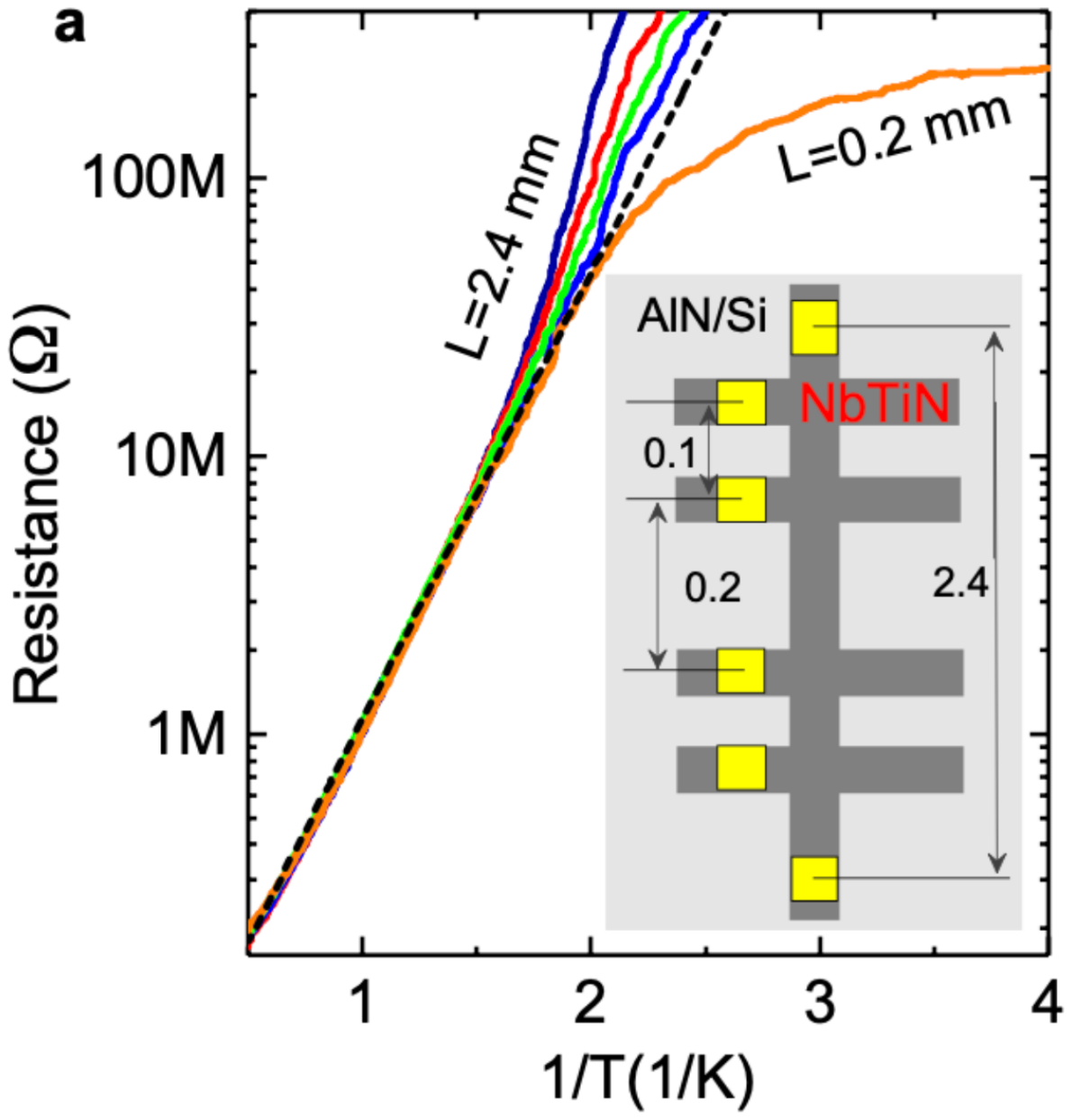}
	\vspace{-2cm}
	\caption{Resistance of NbTin films of different sizes in logarithmic scale as a function of $1/T$. A straight line like the dotted one represents activated behaviour. Large samples show superinsulating hyperactivation of the resistance. When the sample size is sufficiently decreased the string becomes slack, charges are free and the resistance shows the corresponding typical metallic saturation. From \cite{electrostatics}.}
	\label{fig:Fig.10}
\end{figure}

As we have stressed when discussing confinement, the asymptotic freedom of the sine-Gordon model implies a running monopole fugacity and the existence of a dynamically generated scale below which the string tension effectively vanishes. Since the electric interaction is much weaker than the strong interaction, one could expect to directly observe asymptotic freedom by making sufficiently small samples. This is indeed so, as shown in Fig. \ref{fig:Fig.10}. When the sample size is decreased enough, the hyperactivated resistance scaling of the superinsulator goes over directly into a metallic saturation caused by the free charges for which the string is slack on these scales. 

\subsection{Direct measurement of the linear string potential} 
Assume that Cooper pairs $2e$ are bound by a linear potential $U(r) = \sigma r$ with string tension $\sigma$. Is there a universal scaling law that follows from this potential and that can be measured? This is indeed the case. First of all we note that a linear potential corresponds to a constant attractive force $F_{\rm a}$ that can be expressed as $F_{\rm a} = \sigma = 2e V_{\rm c1}/L $ since the lower critical voltage for first passage of current in a sample of size $L$ is $V_{\rm c1} = \sigma L$. Suppose now we apply suddenly at $t=0$ a voltage $V$. The total force on a charge-hole pair is $F_{\rm tot} = F_{\rm r}-F_{\rm a} = 2e\left( V- V_{\rm c1} \right) /L$ and the corresponding centre of mass acceleration for the relative distance is thus $a= (2/m) F_{\rm tot} = (4e/mL) \left( V- V_{\rm c1} \right)$. The equation of motion for the relative distance can be easily solved to give
\begin{equation}
r(t) = {2e\over mL} \left( V- V_{\rm c1} \right) t^2 \ ,
\label{eqmo}
\end{equation}
where we have assumed, for simplicity $r_{\rm initial} = 0$ (the result would not change for a generic initial distance). Current starts to pass when the relative distance reaches the value $L$, i.e. after a shift time 
\begin{equation}
t_{\rm sh} = \sqrt{mL^2\over 2e} \left( V- V_{\rm c1} \right)^{-{1\over 2}}
\label{shti}
\end{equation}
This retardation effect is due to the external voltage having to ``stretch the string" all the way to the boundary and the dynamical exponent $-1/2$ in the scaling of the shift time $t_{\rm sh}$ as a function of reduced voltage $\left( V- V_{\rm c1} \right)$ is characteristic of the linear potential, i.e. of confinement. 

\begin{figure}[t!]
	\includegraphics[width=12cm]{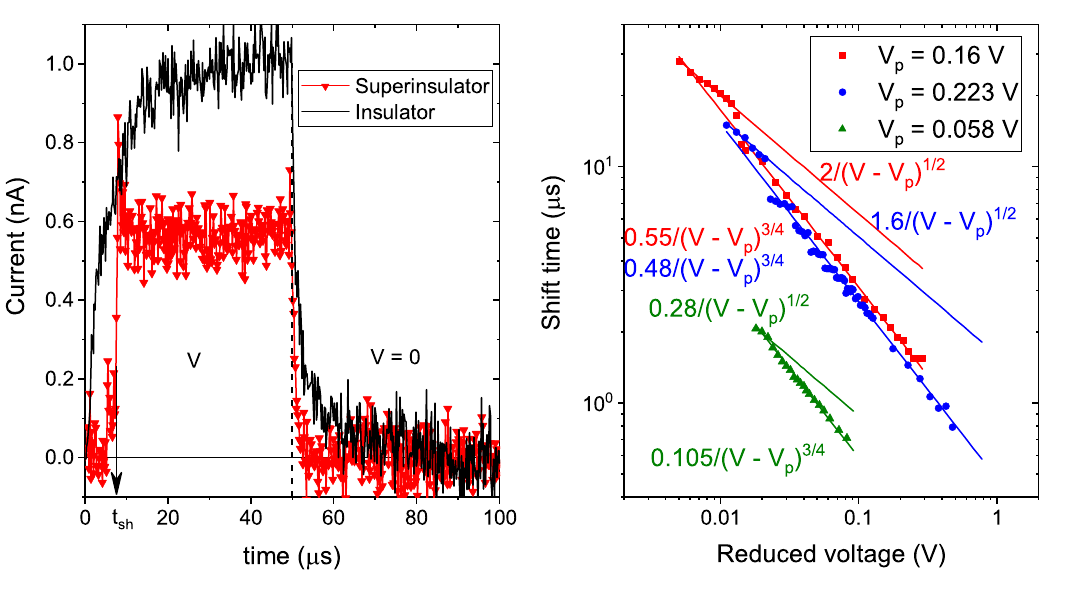}
	%\includegraphics[width=\textwidth]{phase_diagram.pdf}
	%\vspace{-2cm}
	\caption{Left panel: difference in the response to the application of a sudden voltage at $t=0$ between a normal insulator and a superinsulator; note the retardation $t_{\rm sh}$ caused by the string. Right panel: scaling of the shift time $t_{\rm sh}$ with reduced applied voltage; two regimes are clearly identified, the exponent -1/2 corresponds to voltages that stretch the strings to the boundary of the sample, the exponent -3/4 corresponds to higher voltages that break the strings altogehter. From \cite{relaxation}.}
	\label{fig:Fig.11}
\end{figure}

The result of this experiment on three NbTiN superinsulating film samples of different sizes is shown in Fig. \ref{fig:Fig.11}. The dynamic scaling with exponent -1/2 is clearly visible in all three samples. The scaling with exponent -3/4 is characteristic of higher voltages that break the strings altogether, instead of stretching them to the boundary. This experiment is a direct measurement of the linear confining potential in the superinsulating state. 

\section{Why disorder plays no role in the SIT}
A widely held view, rooted in the dirty-boson paradigm \cite{dirtyboson}, attributes the insulating phases observed at the SIT to disorder. In this section we show that this interpretation is incompatible with both the theoretical structure of the problem and the experimental observations.

Two mechanisms dominate the physics of the SIT, 
\begin{itemize}
\item{Topological interactions between charges and vortices.}
\item{Instanton-driven confinement.}
\end{itemize}
Both generate bulk insulating behaviour that is stronger than any form of localization. Disorder becomes at best a subleading effect and does not determine the nature of the phases. 

\subsection{Topological interactions vs. disorder}
The phase-glass scenario for the Bose metal posits that superconducting phases freeze into a disordered landscape, while Cooper pairs retain residual tunnelling through low-energy paths. This assumption, however, neglects the topological mutual statistics of charges and vortices. These topological interactions produce a much stronger form of frustration than disorder. They freeze both charges and vortices into a correlated state in which bulk transport is suppressed. Conduction can persist only along percolative edge channels. 

\subsection{Absence of localizable excitations in the superinsulating phase}
An even more fundamental limitation of disorder-based explanations emerges in the superinsulating phase. Vortices are not Noether charges and can tunnel via instanton events. These proliferate and form a plasma when vortices become tensionless Dirac strings and condense. In the monopole plasma charges are confined by a linear potential, corresponding to flux tubes, or strings, of electric field. Isolated charges are not asymptotic states in the spectrum anymore, consequently there is nothing to be localized. Superinsulation is therefore not a disorder-driven insulator, but a confining phase, qualitatively analogous to confinement in gauge theories and representing an insulating mechanism much stronger than localization.

\subsection{Experimental exclusion of disorder}
The key predictions of this framework are borne out by experiments and are incompatible with disorder-based explanations: 
\begin{itemize}
\item{The Bose-metal phase was observed in highly ordered JJA \cite{van, marcus1, marcus2, marcus3}, demonstrating that disorder is not required for its existence.}
\item{Measurements show that most vortices are frozen, but not all \cite{marcus2}. There are residual mobile ones that account for the observed finite resistance of percolative edge transport.}
\item{Superinsulating phases exhibit BKT scaling in thin films (TiN, NbTiN) \cite{vinokurnature, nbtin} and VFT scaling in thicker films (InO) \cite{shahar}, as expected from confinement physics.}
\item{The same VFT scaling has been observed in ordered nanopatterned TiNb structures \cite{bsi}, excluding disorder as the origin of the effect.}
\item{The characteristic two-kink structure in the I(V) curves \cite{electrostatics} is not reproduced by disorder-based models such as many-body localization, but follows naturally from confinement.}
\item{Finally, the predicted linear interaction potential between Cooper pairs has been measured directly in experiment \cite{relaxation}.}
\end{itemize}
Taken together, these results leave no room for a disorder-based interpretation of the nature of the phases at the SIT.  Disorder can influence quantitative details, such as transition temperatures or couplings, but it does not determine the nature of the phases involved. The SIT is governed, instead by topological interactions and confinement. In particular, the Bose metal is a topologically constrained state, not a phase glass, the insulating phase is a confining phase without localizable excitations, superinsulation. 

Recent measurements in \cite{strunk}, while motivated by the disorder picture, actually provide, instead, further support for the physical picture developed in this review. In ultrathin NbN films, the finite-temperature superconducting transition is shown to retain its BKT character all the way down to the SIT. Close to the critical point, however, the very region of interest, the superfluid stiffness and the BKT temperature fall substantially below the predictions of the disorder-based mean-field description, while the pairing temperature remains finite. Suppression of pairing and thermal vortex fluctuations alone, therefore cannot account for the loss of superconducting coherence. The authors attribute the additional suppression to quantum phase fluctuations, possibly associated with emergent condensate granularity; the simultaneous approach of the low-energy activation scale to the vortex-pair excitation energy further strongly suggests a role for quantum vortex excitations. It can further be shown that a quantum-BKT essential-singularity form \cite{bm, qbkt} describes the near-critical data and the mentioned suppressions comparably to a freely fitted power law and, at the same assumed critical resistance, gives smaller residuals for the three points nearest to the SIT in both observables. Although this limited dataset does not yet uniquely determine the critical law, these findings strongly support the central role of quantum phase fluctuations and emergent granularity emphasized here and lend support to the proposed quantum-vortex interpretation of the critical region.

\subsection{The fata morgana of quantum Griffiths transitions} 

A claim often encountered in the literature is that an apparent divergence of the dynamical exponent $z$ signals that the SIT is a quantum Griffiths transition induced by strong disorder (for a review see, e.g. \cite{wang}). This interpretation is neither unique nor compelling, especially in view of the fact that the SIT is observed in perfectly ordered structures. It has been shown that quantum BKT transitions, arising from a diverging dielectric constant, exhibit the same effective divergence of $z$ \cite{qbkt}. The very phenomenology invoked as evidence for Griffiths physics thus follows directly from quantum BKT criticality. Since this behaviour emerges naturally in superconducting films and in ordered synthetic structures, attributing a divergent $z$ to disorder is unwarranted and, in fact, implausible.

\section{Conclusion}
Superinsulation is the realization of electric confinement in condensed matter systems. It is induced by a condensation of emergent magnetic monopoles and is characterized by the expulsion of electric fields and by an infinite resistance up to a finite deconfinement temperature. As such it is the exact electric-magnetic dual of superconductivity. It has been detected in many systems, including highly ordered ones and the linear confining potential has been confirmed in a direct experiment. This excludes any potential role of disorder, topology and interactions are dominant and superinsulation is a genuine state of matter as is superconductivity. 

\section{Note added in proof}
During proofreading yet another experimental paper appeared \cite{bishop} in which it is shown that, near the SIT, the classical XY description cannot account for the observed suppression of superconductivity, providing further evidence for the central role of quantum phase fluctuations as described in the present review.

\section{Appendix A: the lattice Chern-Simons term} 

\subsection{Two space dimensions}
Here we discuss the proper formulation of the lattice Chern-Simon operator $\epsilon_{\mu \alpha \nu} \partial_{\alpha}$ which respects gauge invariance\,\cite{dst}. We consider a 3D Euclidean lattice with coordinates $\{ x \} = \{ x_0 = vt, x_1, x_2\}$, directions denoted by Greek letters and lattice spacing $\ell$. When not otherwise indicated, a summation is understood over equal Greek letters. We introduce first the forward and backward finite difference and shift operators 
\begin{eqnarray}
d_{\mu} f(x) = {f(x +\ell \hat \mu) -f(x)\over \ell} \ , \qquad S_{\mu } f(x)= f(x +\ell \hat \mu) \ ,
\nonumber \\
\hat d_{\mu} f(x) = {f(x) -f(x -\ell\hat \mu)\over \ell}  \ , \qquad \hat S_{\mu } f(x)= f(x -\ell\hat \mu) \ ,
\label{shift}
\end{eqnarray}
where $\hat \mu$ denotes a unit vector in direction $\mu$. Summation by parts on the lattice interchanges both the two finite differences (with a minus sign) and the two shift operators. Gauge transformations are defined by using the forward finite differences. In terms of these operators one can then define two lattice Chern-Simons terms
\begin{equation}
k_{\mu \nu} = S_{\mu} \epsilon_{\mu \alpha \nu} d_{\alpha} \ , \qquad \qquad \hat k_{\mu \nu} = \epsilon_{\mu \alpha \nu} \hat d_{\alpha} \hat S_{\nu} \ ,
\label{latticecs}
\end{equation}
where no summation is implied over the equal indices $\mu$ and $\nu$. 
Summation by parts on the lattice interchanges also these two operators (without any minus sign). Gauge invariance is then guaranteed by the relations
\begin{equation}
k_{\mu \alpha} d_{\nu} = \hat d_{\mu} k_{\alpha \nu} = 0 \ , \qquad \qquad \hat k_{\mu \nu} d_{\nu} = \hat d_{\mu} \hat k_{\mu \nu} = 0 \ .
\label{gaugeinv}
\end{equation}
The product of the two Chern-Simons terms gives the lattice Maxwell operator
\begin{equation}
k_{\mu \alpha} \hat k_{\alpha \nu} = \hat k_{\mu \alpha} k_{\alpha \nu} = -\delta_{\mu \nu} \Delta_3 +d_{\mu} \hat d_{\nu} \ ,
\label{maxwell}
\end{equation}
where 
\begin{equation}
\Delta_3 = \hat d_{\mu} d_{\mu} ={1\over v^2} \hat d_t d_t + \nabla^2 = {1\over v^2} \hat d_t d_t + \sum_{i=1}^2\hat d_i d_i \ ,
\label{lap}
\end{equation}
is the 3D Laplace operator, while $\nabla^2$ is its 2D counterpart. 

\subsection{Three space dimensions}
Here we extend the results of the previous section to the formulation of a gauge-invariant lattice version of the BF term $\epsilon_{\mu \alpha \nu \rho} \partial_{\alpha} $
in 3D. To this end we introduce the operators
\begin{eqnarray}
	k_{\mu \nu \rho} \equiv S_{\mu }\epsilon _{\mu \alpha
		\nu \rho } d _{\alpha } \ ,
	\nonumber \\
	\hat k_{\mu \nu \rho} \equiv \epsilon _{\mu \nu \alpha \rho}
	\hat d_{\alpha }\hat S_{\rho } \ ,
	\label{kop}
\end{eqnarray}
where $d_{\mu}$, $\hat d_{\mu}$, $S_{\mu}$ and $\hat S_{\mu}$ are the same lattice forward and backward derivatives and shift operators introduced above. The two lattice BF operators are interchanged (no minus sign) upon summation by parts on the lattice, are gauge invariant, 
\begin{eqnarray}
	k_{\mu \nu \rho} d _{\nu } = k_{\mu \nu \rho}
	d _{\rho } = \hat d_{\mu } k_{\mu \nu \rho} = 0 \ ,
	\nonumber \\
	\hat k_{\mu \nu \rho }d _{\rho } = \hat d _{\mu }
	\hat k_{\mu \nu \rho} = \hat d_{\nu }
	\hat k_{\mu \nu \rho } = 0 \ ,
	\label{gaugeinv}
\end{eqnarray}
and satisfy the equations
\begin{eqnarray}
	\hat k_{\mu \nu \rho} k_{\rho \lambda \omega} &&=
	-\left( \delta _{\mu \lambda} \delta_{\nu \omega} - \delta _{\mu \omega}
	\delta_{\nu \lambda } \right) \Delta_4 
	+ \left( \delta _{\mu \lambda }
	d_{\nu } \hat d_{\omega} - \delta _{\nu \lambda } d_{\mu }
	\hat d_{\omega } \right) 
	+ \left( \delta _{\nu \omega} d_{\mu }
	\hat d_{\lambda } - \delta _{\mu \omega} d_{\nu } \hat
	d_{\lambda } \right) \ ,
	\nonumber \\
	\hat k_{\mu \nu \rho} k_{\rho \nu \omega } &&= k_{\mu \nu \rho } \hat
	k_{\rho \nu \omega} = 2 \left( \delta _{\mu \omega } \Delta_4 - d_{\mu }
	\hat d_{\omega } \right) \ ,
	\label{maxwell}
\end{eqnarray}
where $\Delta_4 = \hat d_{\mu} d_{\mu}$ is the 4D  Laplacian.


\begin{thebibliography}{10}
	\expandafter\ifx\csname url\endcsname\relax
	\def\url#1{\texttt{#1}}\fi
	\expandafter\ifx\csname urlprefix\endcsname\relax\def\urlprefix{URL }\fi
	\providecommand{\bibinfo}[2]{#2}
	\providecommand{\eprint}[2][]{\url{#2}}
	
\bibitem{mermin} N. W. Ashcroft and N. D. Mermin, Solid State Physics, Saunders Philadelphia (USA) (1976). 

\bibitem{hasan} M. Z. Hasan and C. L. Kane, Topological insulators, {\it Rev. Mod.  Phys.}  {\bf 82} 3045 (2010). 

\bibitem{mott} N. F. Mott, The basis of the electron theory of metals, with special reference to the transition metals, {\it Proc. Phys. Soc., Sect. A} {\bf 62} 416 (1949).

\bibitem{anderson} P. Anderson, Absence of diffusion in certain random lattices, {\it Phys. Rev.} {\bf 109} 1492-1505 (1958). 

\bibitem{altshuler} D. M. Basko, I. L. Aleiner and B. L. Altshuler, Metal-insulator transition in a weakly interacting many-electron system with localized single-particle states, {\it Ann. Phys.} {\bf 321} 1126-1205 (2006). 

\bibitem{olive} P. Goddard and D. I. Olive, Magnetic monopoles in gauge field theories. {\it Rep. Prog. Phys.} {\bf 41} 91 (1978). 

\bibitem{polyakov}A. M. Polyakov, Compact gauge fields and the infrared catastrophe. {\it Phys. Lett.} {\bf 59}, 82-84 (1975).  

\bibitem{polyakovbook} A.\,M.\,Polyakov, {\it Gauge Fields and Strings}, Harwood Academic Publisher, Chur (Switzerland) (1987).

\bibitem{minnhagen} P.\,Minnhagen, The two-dimensional Coulomb gas, vortex unbinding and superfluid-superconducting films. {\it Rev. Mod. Phys.} {\bf 59}, 1001-1066 (1987). 

\bibitem{greensite}J. Greensite, An introduction to the confinement problem, Springer-Verlag (Berlin) 2011. 

\bibitem{dst} M. C. Diamantini, P. Sodano, C. A. Trugenberger, Gauge theories of Josephson junction arrays. \textit{Nuclear Physics} {\bf B474}, 641-677 (1996). 

\bibitem{vinokurnature}V. M. Vinokur \textit{et~al.} Superinsulator and quantum synchronization. \textit{Nature} \textbf{452}, 613-615 (2008). 

\bibitem{dtv} M. C. Diamantini, C. A. Trugenberger, V. M. Vinokur, Confinement and asymptotic freedom with Cooper pairs. \textit{Comm. Phys.}\,\textbf{1}, 77 (2018). 

\bibitem{coleman} S. Coleman, {\it Aspects of symmetry}. Cambridge University Press, Cambridge (USA) (1985). 

\bibitem{tinkham}M. Tinkahm, Introduction to superconductivity, Dover Publications, New York (USA) (1996). 

\bibitem{goldman1} H. M. Jaeger, D. B. Haviland, B. G. Orr, and A. M. Goldman, Onset of superconductivity in ultrathin granular metal films, {\it Phys. Rev. } {\bf 40} 182 (1989). 

\bibitem{haviland} D. Haviland, Y. Liu and A. Goldman, Onset of superconductivity in the two-dimensional limit, {\it Phys. Rev. Lett.} {\bf 62} 2180-2183 (1989).

\bibitem{hebard} A. Hebard and M. A. Paalanen, Magnetic-field-tuned superconductor-insulator transition in two-dimensional films, {\it Phys. Rev. Lett.} {\bf 65} 927-930 (1990). 

\bibitem{fisher1}M.\,P.\,A.\,Fisher, Quantum\,phase\,transitions\,in\,disordered\,two-dimensional\,superconductors. {\it Phys.\,Rev.\,Lett.}\,{\bf 65},\,923-926 (1990).

\bibitem{fisher2} M.\,P.\,A.\,Fisher, G.\,Grinstein, S.\,M.\,Girvin, Presence of quantum diffusion in two dimensions: Universal resistance at the superconductor-insulator transition. \textit{Phys. Rev. Lett.} \textbf{64}, 587-590 (1990).

\bibitem{goldmanrev} 
A.\,M.\,Goldman, Superconductor-insulator transitions. \textit{Int. J. Mod. Phys}.\,B\,\textbf{24}, 4081--4101 (2010).

\bibitem{phillipsrev1} P. Phillips, Not just a phase, {\it Nature Physics} {\bf 12} 206-207 (2016). 

%\bibitem{phillipsrev2} P. Phillips, Free at last: Bose metal uncaged, {\it Science} {\bf 366} 1450-1451 (2019). 

\bibitem{boserev} M. C. Diamantini and C. A. Trugenberger, Bose metals, from prediction to realization, {\it Materials} {\bf 17} 4924 (2024). 

\bibitem{dst2}M. C. Diamantini, P. Sodano and C. A. Trugenberger, Self duality and oblique confinement in planar gauge theories, {\it Nucl. Phys.} {\bf B448} 505-532 (1995). 

\bibitem{moncond} M. C. Diamantini, C. A. Trugenberger, V. M. Vinokur, Quantum magnetic monopole condensate, {\it Nature Comm. Phys. } {\bf 4} 25 (2021). 

\bibitem{van} H. S. J. van der Zant, W. J. Elion, L. J. Geerlings and J. E. Mooji, Quantum phase transitions in two dimensions: Experiments in Josephson-junction arrays, {\it Phys. Rev.} {\bf B54} 10081 (1996). 

\bibitem{clean1} A. W. Tsen et al., Nature of the quantum metal in a two-dimensional crystalline superconductor, {\it Nature Physics} {\bf 12} 208 (2015). 

\bibitem{clean2} Y. Saito, Y. Kasahara, J. Ye, Y. Iwasa and T. Nojima, Metallic ground state in an ion-gated two-dimensional superconductor, {\it Science} {\bf 350} 409-413 (2015). 

\bibitem{marcus1} C. G. L. Bottcher {\it et. al.} Superconducting, insulating and anomalous metallic regimes in a gated two-dimensional semiconductor-superconductor array. {\it Nat. Phys.} {\bf 14}, 1138-1144 (2018).

\bibitem{yang} C. Yang et al., Intermediate bosonic metallic state in the superconductor-insulator transition, {\it Science} {\bf 366} 1505 (2019). 

\bibitem{kapitulnik} X. Zhang, A. Palevski and A. Kapitulnik, Anomalous metals: From ``failed superconductor” to ``failed insulator”, {\it PNAS} {\bf  119} e2202496119 (2022).

\bibitem{marcus2} C. G. L. Bottcher {\it et. al.}, Dynamical vortex transitions in a gate-tunable, two-dimensional Josephson junction array, {\it Phys. Rev.} {\bf B108} 134517 (2023).  

\bibitem{marcus3} C. G. L. Bottcher {\it et. al.} The Berezinskii-Kosterlitz-Thouless transition and anomalous metallic phase in a hybrid Josephson junction array, {\it Phys. Rev. } {\bf B110} L180502 (2024). 

\bibitem{shahar} G. Sambandamurthy, L. W. Engels, A. Johansson, E. Peled \& D. Shahar, Experimental evidence for a collective insulating state in two-dimensional superconductors, {\it Phys. Rev. Lett.} {\bf 94}, 017003 (2005). 

\bibitem{nbtin} A. Yu. Mironov et al. Charge Berezinskii-Kosterlitz-Thouless transition in superconducting NbTiN films, {\it Scientific Report} {\bf 8} 4082 (2018).

\bibitem{electrostatics}M. C. Diamantini, L. Gammaitoni, C. Strunk, S. V. Postolova, A. Yu. Mironov, C. A. Trugenberger and V. M. Vinokur, Direct probe of the interior of an electric pion in a Cooper pair superinsulator. {\it Nat. Comm. Phys.} {\bf 3}, 142 (2020). 

\bibitem{relaxation} A. Yu. Mironov et al. Relaxation electrodynamics of superinsulators. {\it Scientific Reports} {\bf 12} 19918 (2022). 

\bibitem{bsi} A. Yu. Mironov, C. A. Trugenberger, M. C. Diamantini, D. A. Nasimov and V. M. Vinokur, Bulk superinsulation and polar nematic orders in nanopatterned NbTiN, {\it Nature Comms.} {\bf 16} 4395 (2025). 

\bibitem{efros1} A. L. Efros and B. I. Shklovskii, Critical behaviour of conductivity and dielectric constant near the metal-non-metal transition threshold. 
{\it Phys. Status Solidi B Basic Res.} {\bf 76} 475-485 (1976). 	

\bibitem{efros2} A. L. Efros and B. I. Shklovskii, Electronic Properties of Doped Semiconductors, Springer-Verlag, Berlin (1984).

\bibitem{baturina} T. Baturina and V. M. Vinokur, Superinsulator-Superconductor Duality in Two Dimensions, {\it Ann. Phys.} {\bf 331} 236-257 (2013). 

\bibitem{bakkali} H. Bakkali et. al. Divergence of the dielectric constant in ultrathin granular metal films near the percolation threshold, {\it New. J. Phys.} {\bf 22} 083018 (2020). 

\bibitem{chaplik}A. V. Chaplik and M. V. Entin, Charged impurities in very thin layers, {\it Zh. Eksp. Teor. Fiz.} {\bf 61} 2496-2503 (1971).

\bibitem{keldysh} L. V. Keldysh, Coulomb interaction in thin semiconductor and semimetal films, {\it JETP Lett.} {\bf 29} 658-661 (1979). 
	
\bibitem{jackiw} R. Jackiw and S. Templeton, How super-renormalizable interactions cure infrared divergences. {\it Phys. Rev.}{\bf D23}, 2291 (1981). 
	
\bibitem{super2D} M. C. Diamantini, C. A. Trugenberger and V. M. Vinokur, How planar superconductors cure their infrared divergences, {\it JHEP} {\bf 10} 100 (2022). 

\bibitem{typeIII} M. C. Diamantini, C. A. Trugenberger, S.-Z. Chen, Y.-J. Lu, C.-T. Liang and V. M. Vinokur, Type-III superconductivity, {\it Adv. Sci.} 2206523 (2023). 

\bibitem{typeIIIvort} M. C. Diamantini, C. A. Trugenberger and V. M. Vinokur, Topological gauge theory of vortices in type-III superconductors, {\it Phys. Rev.} {\bf B110} 094506 (2024). 

\bibitem{sacepe1} B. Sac\'ep\'e \textit{et al}. Disorder-induced inhomogeneities of the superconducting state close to the superconductor-insulator transition. {\it Phys. Rev. Lett.} {\bf 101} 157006 (2008). 

\bibitem{granular} D. Kowal \& Z. Ovadyahu, Disorder induced granularity in an amorphous superconductor. {\it Solid St. Comm.} {\bf 90}, 783-786 (1994).

\bibitem{aharonovbohm} Y.\,Aharonov \& D.\,Bohm, Significance of electromagnetic potentials in quantum theory. {\it Phys. Rev. } {\bf 115}, 485--491 (1961).

\bibitem{aharonovcasher}Y.\,Aharonov, A.\,Casher, Topological quantum effects for neutral particles. \textit{Phys. Rev. Lett}. \textbf{53}, 319--321 (1984).

\bibitem{negele} J. W. Negele and H. Orland, Quantum many-particle systems, World Scientific, Singapore (1988).

\bibitem{wu} Y.-S. Wu, General theory for quantum statistics in two dimensions, {\it Phys. Rev. Lett.} {\bf 52} 2103 (1984).

\bibitem{kaufmann}L.\,H.\,Kaufmann, {\it Formal knot theory}, Princeton University Press, Princeton (1983). 

\bibitem{wilczek}F.\,Wilczek, Disassembling anyons. {\it Phys. Rev. Lett.} {\bf 69}, 132--135 (1992). 

\bibitem{wilczekbook}F.\,Wilczek, Fractional statistics and anyon superconductivity. World Scientific, Singapore (1990). 

\bibitem{jackiw2} S.\,Deser, R.\,Jackiw \& S.\,Templeton, Three-dimensional massive gauge theories.
 {\it Phys. Rev. Lett.} {\bf 48}, 975 (1982).

\bibitem{bti1} Y.-M.\,Lu,  A.\,Vishwanath, Theory and Classification of interacting integer topological phases in two dimensions: a Chern-Simons approach. {\it Phys. Rev} B\,{\bf 86}, 125119 (2012); 

\bibitem{bti2} C.\,Yang and T.\,Senthil, Boson topological insulators: A window into highly entangled quantum phases.
\textit{Phys. Rev}. B\,\textbf{87}, 235122 (2013); 

\bibitem{bti3} X. Chen, Z.-C. Gu, Z.-X. Liu \& X.-G. Wen, Symmetry protected topological orders and the group cohomology of their symmetry group. {\it Phys. Rev. } B\,{\bf 87}, 155114 (2013). 

\bibitem{djt}G. Dunne, R. Jackiw and C. A. Trugenberger, Chern-Simons theory in the Schr\"odinger representation, {\it Ann. Phys.} {\it 194} 197-223 (1989). 

\bibitem{bm} M. C. Diamantini, A. Yu. Mironov, S..V. Postolova,  X. Liu, Z. Hao, D. M. Silevitch, Ya. Kopelevich, P. Kim, C. A. Trugenberger and V. M. Vinokur, Bosonic topological intermediate state in the superconductor-insulator transition. {\it Phys. Lett.} {\bf A384}, 126570 (2020). 

\bibitem{phillips3} D. Dalidovich and P. Phillips, Phase Glass is a Bose Metal: A New Conducting State in Two Dimensions, {\it Phys. Rev. Lett.} {\bf 89} 027001 (2002). 

\bibitem{phillips4} P. Phillips, The elusive Bose metal, {\it Science} {\bf 302} 243-247 (2003). 

\bibitem{nodisorder}
M.\,C.\,Diamantini, C.\,A.\,Trugenberger, V.\,M.\,Vinokur, The superconductor-insulator transition in absence of disorder,  {\it Phys. Rev.} {\bf B103}, 174516 (2021). 

\bibitem{wen} Y.-G. Wen, Theory of the edge states in fractional quantum Hall effects, {\it Int. J. Mod. Phys.} {\bf B6} 1711-1762 (1992). 

\bibitem{qps1}D. S. Golubev and A. D. Zaikin, Quantum tunnelling of the order parameter in superconducting nanowires, {\it Phys. Rev.} {\bf B64} 014504 (2001).

\bibitem{qps2}K. Yu. Arutyunov, D. S. Golubev and A. D. Zaikin, Superconductivity in one dimension. {\it Phys. Rep.} {\bf 464} 1-70 (2008). 

\bibitem{das} D. Das and S. Doniach, Existence of a Bose metal at $T=0$, {\it Phys. Rev. } {\bf B60} 1261 (1999). 

\bibitem{cc} J. T. Chalker and P. D. Coddington, Percolation, quantum tunnelling and the integer Hall effect, {\it J. Phys.} {\bf C21} 2665 (1988). 

\bibitem{jjarev} R. Fazio \& H. van der Zant, Quantum phase transitions and vortex dynamics in superconducting netrworks. {\it Phys. Rep.} {\bf 355}, 235-334 (2001). 

%\bibitem{marcus4}C. G. L Bottcher, F. Nichele, J. Shabani, C. J. Palmstrøm and C. M Marcus, Dynamical vortex transitions in a gate-tunable, two-dimensional Josephson junction array, {\it Phys. Rev. } {\bf B108} 134517 (2023). 

\bibitem{fazio}R.\,Fazio \& G.\,Sch\"on, Charge and vortex dynamics in arrays of tunnel junctions. {\it Phys. Rev.}\,B\,{\bf 43} 5307-5320 (1991).

\bibitem{hansson}T. H. Hansson, V. Oganesyan and S. L. Sondhi, Superconductors are topologically ordered, {\it Ann. Phys.} {\bf 313} 497 (2004). 

\bibitem{higgsless} M. C. Diamantini and C. A. Trugenberger, Higgsless superconductivity from topological defects in compact BF terms, {\it Nucl. Phys.} {\bf 891} 401-419 (2015).

\bibitem{ber} V. L. Berezinskii, Destruction of long-range order in one-dimensional and two-dimensional
systems having a continuous symmetry group I. Classical systems. \textit{Sov. Phys.--JETP} \textbf{32}, 493--500 (1970).

\bibitem{kt1} J. M. Kosterlitz, D. J. Thouless, Long range order and metastability in two dimensional solids and superfluids. (Application of dislocation theory). \textit{Journal of Physics C: Solid State Physics,} \textbf{5}, L124 (1972).

\bibitem{kt2}J. M. Kosterlitz, D. J. Thouless, Ordering, metastability and phase transitions in two-dimensioal systems. \textit{J. Phys. C: Solid State Phys.}  \textbf{6}, 1181--1203 (1973).

\bibitem{qbkt}M. C. Diamantini, C. A. Trugenberger, V. M. Vinokur, Berezinskii–Kosterlitz–Thouless quantum transition in two dimensions, {\it Materials} {\bf 19} 868 (2026). 

\bibitem{mussardo} G. Mussardo, Statistical field theory, Oxford University Press, Oxford (2020). 

\bibitem{poccia} C. A Trugenberger, M. C. Diamantini, N. Poccia, F. S. Nogueira \& V. M. Vinokur, Magnetic monopoles and superinsulation in Josephson junction arrays. {\it Quantum Reports} {\bf 2} 388-399 (2020). 

\bibitem{kleinertbook} H. Kleinert, Gauge Fields in Condensed Matter, World Scientific, Singapore (1989). 

\bibitem{parra}C. Parra, F. Niemstemski, A. W. Contryman, P. Giraldo-Gallo, T. H. Geballe, I. R. Fisher and H. C. Manoharan, Signatures of two-dimensional superconductivity emerging within a three-dimensional host superconductor, {\it PNAS} {\bf 118}, e2017810118 (2021).

\bibitem{bowick}T. Allen, M. Bowick and A. Lahiri, Topological Mass Generation in 3+1 dimensions. {\it Mod. Phys. Lett.} A\,{\bf 6},  559-571 (1991).

\bibitem{nelson} Nelson,\,D., Piran,\,T., \& Weinberg,\,S. eds. {\it Statistical Mechanics of Membranes and Surfaces}, World Scientific, Singapore (2004). 

\bibitem{vft} M.\,C.\,Diamantini, L.\,Gammaitoni, C.\,A.\,Trugenberger and V.\,M.\,Vinokur, Vogel-Fulcher-Tamman criticality of 3D superinsulators,  {\it Scientific Reports}\,{\bf 8}, 15718 (2018). 

\bibitem{dirtyboson}M. P. A. Fisher, P. B. Weichman, G. Grinstein and D. S. Fisher, Boson localization and the superfluid-insulator transition, {\it Phys. Rev. } {\bf B40} 546 (1989). 

\bibitem{strunk} A. Weitzel et al., Origin of the superconductor-insulator transition in disordered two-dimensional films, arXiv:2608.07101. 

\bibitem{wang} Z. Wang, Y. Liu, C. Ji and J. Wang, Quantum phase transitions in two-dimensional superconductors: a review on recent experimental progress, {\it Rep. Prog. Phys.} {\bf 87} 014502 (2024). 

\bibitem{bishop} L. Bishop-van Horn et al., Imaging how fluctuations destroy superconductivity in two dimensions, arXiv:2609.28837. 




\iffalse



	
\bibitem{negele} J. W. Negele and H. Orland, Quantum many-particle systems, World Scientific, Singapore (1988).

%\bibitem{feigelman} I. Poboiko and M. Feigel'man, Mean-field theory of first-order quantum superconductor-insulator transition, {\it Sci. Post. Phys.} {\bf 17} 066 (2024). 
	
\bibitem{efros1} A. L. Efros and B. I. Shklovskii, Critical behaviour of conductivity and dielectric constant near the metal-non-metal transition threshold. 
{\it Phys. Status Solidi B Basic Res.} {\bf 76} 475-485 (1976). 	

\bibitem{efros2} A. L. Efros and B. I. Shklovskii, Electronic Properties of Doped Semiconductors, Springer-Verlag, Berlin (1984).
	

%\bibitem{feigelmanrev} B.\,Sac\'ep\'e, M.\,Feigel'man and T.\,M.\,Klapwijk, Quantum breakdown of superconductivity in low-dimensional materials, {\it Nature Physics} {\bf 16} 734--746 (2020). 

\bibitem{baturina} T. Baturina and V. M. Vinokur, Superinsulator-Superconductor Duality in Two Dimensions, {\it Ann. Phys.} {\bf 331} 236-257 (2013). 

\bibitem{bakkali} H. Bakkali et. al. Divergence of the dielectric constant in ultrathin granular metal films near the percolation threshold, {\it New. J. Phys.} {\bf 22} 083018 (2020). 

\bibitem{vojta} T. Vojta, Phases and phase transitions in disordered quantum systems, arXiv 1301.7746 (2013). 


\bibitem{chaplik}A. V. Chaplik and M. V. Entin, Charged impurities in very thin layers, {\it Zh. Eksp. Teor. Fiz.} {\bf 61} 2496-2503 (1971).

\bibitem{keldysh} L. V. Keldysh, Coulomb interaction in thin semiconductor and semimetal films, {\it JETP Lett.} {\bf 29} 658-661 (1979). 
	
\bibitem{jackiw} R. Jackiw and S. Templeton, How super-renormalizable interactions cure infrared divergences. {\it Phys. Rev.}{\bf D23}, 2291 (1981). 
	
\bibitem{super2D} M. C. Diamantini, C. A. Trugenberger and V. M. Vinokur, How planar superconductors cure their infrared divergences, {\it JHEP} {\bf 10} 100 (2022). 

\bibitem{typeIII} M. C. Diamantini, C. A. Trugenberger and V. M. Vinokur, Type-III superconductivity, {\it Adv. Sci.} 2206523 (2023). 

\bibitem{typeIIIvort} M. C. Diamantini, C. A. Trugenberger and V. M. Vinokur, Topological gauge theory of vortices in type-III superconductors, {\it Phys. Rev.} {\bf B110} 094506 (2024). 

\bibitem{sacepe1} B. Sac\'ep\'e \textit{et al}. Disorder-induced inhomogeneities of the superconducting state close to the superconductor-insulator transition. {\it Phys. Rev. Lett.} {\bf 101} 157006 (2008). 

\bibitem{nbtin} A. Yu. Mironov et al. Charge Berezinskii-Kosterlitz-Thouless transition in superconducting NbTiN films, {\it Scientific Report} {\bf 8} 4082 (2018). 

\bibitem{granular} D. Kowal \& Z. Ovadyahu, Disorder induced granularity in an amorphous superconductor. {\it Solid St. Comm.} {\bf 90}, 783-786 (1994).

\bibitem{bm} M. C. Diamantini, A. Yu. Mironov, S..V. Postolova,  X. Liu, Z. Hao, D. M. Silevitch, Ya. Kopelevich, P. Kim, C. A. Trugenberger and V. M. Vinokur, Bosonic topological intermediate state in the superconductor-insulator transition. {\it Phys. Lett.} {\bf A384}, 126570 (2020). 

\bibitem{baiguera} S. Baiguera, Aspects of non-relativistic quantum field theories, {\it Eur. Phys. J.} {\bf C84} 268 (2024). 

\bibitem{xing} Y. Xing et al., Quantum Griffiths singularity of superconductor-metal transition in Ga thin films. {\it Science} {\bf 350} 542-545 (2015). 

\bibitem{zhang} C. Zhang et al., Quantum Griffiths singularities in TiO superconducting thin films with insulating normal states. {\it npg Asia Materials} {\bf 11} 73 (2019). 

\bibitem{jing} T.-Y. Jing et al., Quantum phase transition in NbN superconducting thin films, {\it Phys. Rev.} {\bf B107} 184515 (2023). 

\bibitem{yadav} S. Yadav, M. P. Saravanan and Sangeeta Sahoo, Emergence of quantum Griffiths singularity in disordered TiN thin films. {\it Communications Physics} {\bf 7} 215 (2024).

\bibitem{anderson} P. Anderson, Absence of diffusion in certain random lattices, {\it Phys. Rev.} {\bf 109} 1492-1505 (1958). 

\bibitem{abrahams} A. Abrahams, Fifty years of Anderson localization, World Scientific, Singapore (2010). 
	
\bibitem{goldman1} H. M. Jaeger, D. B. Haviland, B. G. Orr, and A. M. Goldman, Onset of superconductivity in ultrathin granular metal films, {\it Phys. Rev. } {\bf 40} 182 (1989). 

\bibitem{haviland} D. Haviland, Y. Liu and A. Goldman, Onset of superconductivity in the two-dimensional limit, {\it Phys. Rev. Lett.} {\bf 62} 2180-2183 (1989).

\bibitem{hebard} A. Hebard and M. A. Paalanen, Magnetic-field-tuned superconductor-insulator transition in two-dimensional films, {\it Phys. Rev. Lett.} {\bf 65} 927-930 (1990). 

\bibitem{goldmanrev} 
A.\,M.\,Goldman, Superconductor-insulator transitions. \textit{Int. J. Mod. Phys}.\,B\,\textbf{24}, 4081--4101 (2010).

\bibitem{feigelman} B.\,Sac\'ep\'e, M.\,Feigel'man and T.\,M.\,Klapwijk, Quantum breakdown of superconductivity in low-dimensional materials, {\it Nature Physics} {\bf 16} 734--746 (2020). 

\bibitem{van} H. S. J. van der Zant, W. J. Elion, L. J. Geerlings and J. E. Mooji, Quantum phase transitions in two dimensions: Experiments in Josephson-junction arrays, {\it Phys. Rev.} {\bf B54} 10081 (1996). 

\bibitem{das} D. Das and S. Doniach, Existence of a Bose metal at $T=0$, {\it Phys. Rev. } {\bf B60} 1261 (1999). 

\bibitem{mason} N. Mason and A. Kapitulnik, Dissipation effects on the superconductor-insulator transition in 2-D superconductors, {\it Phys. Rev. Lett.} {\bf 82} 5341 (1999). 



\bibitem{shahar} I. Tamir et al., Sensitivity of the superconducting state in thin films, {\it Sci. Adv.} {\bf 5} eeau3826 (2019). 



\bibitem{yang} C. Yang et al., Intermediate bosonic metallic state in the superconductor-insulator transition, {\it Science} {\bf 366} 1505 (2019). 
 

\bibitem{sacepe2}B. Sac\'ep\'e et al., Localization of preformed Cooper pairs in disordered superconductors. {\it Nature Physics} {\bf 7}, 239-244 (2011). 


\bibitem{fazio}R.\,Fazio \& G.\,Sch\"on, Charge and vortex dynamics in arrays of tunnel junctions. {\it Phys. Rev.}\,B\,{\bf 43} 5307--5320 (1991).

\bibitem{jjarev} R. Fazio \& H. van der Zant, Quantum phase transitions and vortex dynamics in superconducting netrworks. {\it Phys. Rep.} {\bf 355}, 235-334 (2001). 

\bibitem{planar} M. C. Diamantini, C. A. Trugenberger and V. M. Vinokur, How planar superconductors cure their infrared divergences, {\it JHEP} {\bf 10} 100 (2022). 

\bibitem{phillips3} D. Dalidovich and P. Phillips, Phase Glass is a Bose Metal: A New Conducting State in Two Dimensions, {\it Phys. Rev. Lett.} {\bf 89} 027001 (2002). 

\bibitem{phillips4} P. Phillips, The elusive Bose metal, {\it Science} {\bf 302} 243-247 (2003). 

\bibitem{carruthers} P. Carruthers and M. M. Nieto, Phase and angle variables in quantum mechanics, {\it Rev. Mod. Phys.} {\bf 40} 411 (1968). 

\bibitem{karstrup} H. A. Karstrup, Quantization of the optical phase space $S^2 = {phi mod 2pi, I > 0}$ in Terms of the Group SO(1,2), {\it Fort. Phys.} {\bf 51} 975-1134 (2006).

\bibitem{rocca} F. Rocca and M. Sirugue, Phase operator and condensed systems, {\it Comm. Math. Phys.} {\bf 34} 111-121 (1973). 

\bibitem{clean1} A. W. Tsen at al., {\it Nature Physics} {\bf 12} 208 (2015). 

\bibitem{clean2} Y. Saito, Y. Kasahara, J. Ye, Y. Iwasa and T. Nojima, Metallic ground state in an ion-gated two-dimensional superconductor, {\it Science} {\bf 350} 409-413 (2015). 

\bibitem{marcus1} C. G. L. Bottcher {\it et. al.} Superconducting, insulating and anomalous metallic regimes in a gated two-dimensional semiconductor-superconductor array. {\it Nat. Phys.} {\bf 14}, 1138-1144 (2018).

\bibitem{marcus2} C. G. L. Bottcher {\it et. al.} The Berezinskii-Kosterlitz-Thouless transition and anomalous metallic phase in a hybrid Josephson junction array, arXiv:2210.00318. 

\bibitem{marcus3} C. G. L. Bottcher {\it et. al.}, Dynamical vortex transitions in a gate-tunable, two-dimensional Josephson junction array, {\it Phys. Rev.} {\bf B108} 134517 (2023).  

\bibitem{geom} A. Hegg, J. Hou and W. Ku, Geometric frustration produces long-sought Bose metal phase of quantum matter, {\it PNAS} {\bf 118} e2100545118 (2021). 

%\bibitem{phase}F. Rocca and M. Sirugue, {\it Comm. Math. Phys.} {\bf 34} 111-121 (1973). 

\bibitem{book}C. A. Trugenberger, Superinsulators, Bose metals and high-$T_c$ superconductors: the quantum physics of emergent magnetic monopoles, World Scientific, Singapore (2022). 

\bibitem{tinkham} M.\,Tinkham, {\it Introduction to superconductivity} Dover Publications, New York (1996). 

	
\bibitem{golubev}D. S. Golubev and A. D. Zaikin, Quantum tunnelling of the order parameter in superconducting nanowires, {\it Phys. Rev.} {\bf B64} 014504 (2001).

\bibitem{arutyunov}K. Yu. Arutyunov, D. S. Golubev and A. D. Zaikin, Superconductivity in one dimension. {\it Phys. Rep.} {\bf 464} 1-70 (2008). 

\bibitem{aharonovbohm} Y.\,Aharonov \& D.\,Bohm, Significance of electromagnetic potentials in quantum theory. {\it Phys. Rev. } {\bf 115}, 485--491 (1961).

\bibitem{aharonovcasher}Y.\,Aharonov, A.\,Casher, Topological quantum effects for neutral particles. \textit{Phys. Rev. Lett}. \textbf{53}, 319--321 (1984).

\bibitem{wilczek}F.\,Wilczek, Disassembling anyons. {\it Phys. Rev. Lett.} {\bf 69}, 132--135 (1992). 

\bibitem{wu} Y.-S. Wu, General theory for quantum statistics in two dimensions, {\it Phys. Rev. Lett.} {\bf 52} 2103 (1984).

\bibitem{kaufmann}L.\,H.\,Kaufmann, {\it Formal knot theory}, Princeton University Press, Princeton (1983). 

\bibitem{jackiw2} S.\,Deser, R.\,Jackiw \& S.\,Templeton, Three-dimensional massive gauge theories.
 {\it Phys. Rev. Lett.} {\bf 48}, 975 (1982).

\bibitem{bti1} Y.-M.\,Lu,  A.\,Vishwanath, Theory and Classification of interacting integer topological phases in two dimensions: a Chern-Simons approach. {\it Phys. Rev} B\,{\bf 86}, 125119 (2012); 

\bibitem{bti2} C.\,Yang and T.\,Senthil, Boson topological insulators: A window into highly entangled quantum phases.
\textit{Phys. Rev}. B\,\textbf{87}, 235122 (2013); 

\bibitem{bti3} X. Chen, Z.-C. Gu, Z.-X. Liu \& X.-G. Wen, Symmetry protected topological orders and the group cohomology of their symmetry group. {\it Phys. Rev. } B\,{\bf 87}, 155114 (2013). 

\bibitem{jackiwbook} R. Jackiw, ``Topological investigations of quantized gauge theories", in
S. B. Treiman, R. Jackiw, B. Zumino, E. Witten, {\it Current algebra and anomalies} World Scientifc, Singapore (1985).

\bibitem{djt}G. Dunne, R. Jackiw and C. A. Trugenberger, Chern-Simons theory in the Schr\"odinger representation, {\it Ann. Phys.} {\it 194} 197-223 (1989). 

\bibitem{wen} Y.-G. Wen, Theory of the edge states in fractional quantum Hall effects, {\it Int. J. Mod. Phys.} {\bf B6} 1711-1762 (1992). 

\bibitem{cc} J. T. Chalker and P. D. Coddington, Percolation, quantum tunnelling and the integer Hall effect, {\it J. Phys.} {\bf C21} 2665 (1988). 


%\bibitem{marcus4} C. G. L. Bottcher {\it et. al.}, Dynamical vortex transitions in a gate-tunable Josephson junction array, arXiv:2212.0865. 

%\bibitem{wenzee} X. G. Wen and A. Zee, Classification of Abelian quantum Hall states and matrix formulation of topological fluids. {\it Phys. Rev. } {\bf B46}, 2290 (1992). 

\bibitem{nodisorder}
M.\,C.\,Diamantini, C.\,A.\,Trugenberger, V.\,M.\,Vinokur, The superconductor-insulator transition in absence of disorder,  {\it Phys. Rev.} {\bf B103}, 174516 (2021). 

\bibitem{wilczekbook}F.\,Wilczek, Fractional statistics and anyon superconductivity. World Scientific, Singapore (1990). 

\bibitem{magnus} E.\,B.\,Sonin, The Magnus force in superfluids and superconductors. {\it Phys. Rev.} {\bf B55}, 485 (1996). 

\bibitem{senthillevin} T.\,Senthil and M.\,Levin, Integer\,quantum\,Hall\,effect\,for\,bosons, {\it 
Phys.\,Rev.\,Lett.} {\bf 110}, 046801 (2013). 

\bibitem{xu}
C.\,Xu and T.\,Senthil, Wave functions of bosonic symmetry protected topological phases. {\it 
Phys. Rev.}  {\bf B87}  174412 (2013). 

\bibitem{kivelson} K.-S- Kim and S. A. Kivelson, The quantum Hall effect in the absence of disorder, {\it Quantum Mater.} {\bf 6} 22 (2021). 







%\bibitem{Goldman2013}
%A.\,M.\,Goldman, The Berezinskii-Kosterlitz-Thouless Transition in Superconductors, in: \textit{40 YEARS OF BEREZINSKII-KOSTERLITZ-THOULESS THEORY, Ed Jorge Jos\'{e}}. World Scientific Publishing Co, Singapore, 2013, pp. 135 -- 160.
	


%\bibitem{dtv2019}
%M.\,C.\,Diamantini, C.\,A.\,Trugenberger, and V.\,M.\,Vinokur,  Topological gauge theory of the superconductor-insulator transition, in: \textit{Topological Phase Transitions and New Developments}, World Scientific, 2019, pp.197--221.
	
	
%\bibitem{larkin} A. Larkin and A. Varlamov, Theory of fluctuations in superconductors, Clarendon Press, Oxford (2005).  	
	
%\bibitem{palle} G. Palle and D. K. Sunko Physical limitations of the Hohenberg-Mermin-Wagner	theorem, {\it J. Phys A: Math. Theor} {\bf 54} 315001 (2021). 



%\bibitem{baturina} Cooper pairs outside SC

%\bibitem{topsc}
%M.\,C.\,Diamantini, P.\,Sodano and C.\,A.\,Trugenberger, Superconductors with topological order, {\it Eur. Phys. J.}\,B\,{\bf 53}, 19 (2016).

%\bibitem{higgsless} 
%M.\,C.\,Diamantini and C.\,A.\,Trugenberger, Higgsless superconductivity from topological defects in compact BF terms. {\it Nucl. Phys.} {\bf 891} 401--419 (2015).




%\bibitem{zarembo} N.\,O.\,Agasian and K.\,Zarembo, Phase structure and non-perturbative states in three-dimensional adjoint Higgs model. {\it Phys. Rev.}\,D\,{\bf 57}, 2475--2485 (1998). 
	

 

 %\bibitem{dtv1}
%M. C. Diamantini, C. A. Trugenberger, V. M. Vinokur, Confinement and asymptotic freedom with Cooper pairs. \textit{Comm. Phys.}\,\textbf{1}, 77 (2018). 

%\bibitem{dtv2}
%M.\,C.\,Diamantini, L.\,Gammaitoni, C.\,A.\,Trugenberger and V.\,M.\,Vinokur, Vogel-Fulcher-Tamman criticality of 3D superinsulators,  {\it Scientific Reports}\,{\bf 8}, 15718 (2018). 

%\bibitem{dtv3}
%M.\,C.\,Diamantini\,\textit{et\,al}.\,Direct\,probe\,of\,the\,interior\,of\,an\,electric pion\,in\,a\,Cooper\,pair\,superinsulator,\,\textit{Comm.\,Phys.}\,\textbf{3},\,142\,(2020).

%\bibitem{Likharev1979}
%K.\,K.\,Likharev, Superconducting weak links. \textit{Reviews of Modern Physics}\,\textbf{51}, 101--115 (1979). 

%\bibitem{fisher}M.\,P.\,A.\,Fisher \& G.\,Grinstein, Quantum critical phenomena in charged superconductors. {\it Phys. Rev. Lett.} {\bf 60}, 208--211 (1988).

%\bibitem{girvin} S.\,M.\,Girvin, M.\,Wallin, M.-C.\,Cha, M.\,P.\,A.\,Fisher \& P.\,Young, Universal conductivity as the superconductor-insulator transition in two dimensions. {\it Prog. Theor. Phys. Supp.} {\bf 107}, 135-144 (1992). 

%\bibitem{ballistic1}A.\,van\,Otterlo, R.\,Fazio \& G.\,Sch\"on, Quantum vortex dynamics in Josephson junction arrays. {\it Physica}\,B\,{\bf 203}, 504--512 (1994). 

%\bibitem{ballistic2}H. van der Zant, F. C. Fritschy, T. P. Orlando \& J. E. Mooji, Ballistic motion of vortices in Josephson junction arrays. {\it Europhys. Lett.} {\bf 18}, 343-512 (1992). 

%\bibitem{csqm}G.\,Dunne,\,R.\,Jackiw,\,and\,C.\,A.\,Trugenberger,\,Topological\,(Chern-Simons) quantum mechanics, {\it Phys. Rev.}\,D\,{\bf 41}, 661--666 (1990).

%\bibitem{kogut} T.\,Banks,\,R.\,Myerson,\,and\,J.\,Kogut,\,Phase\,transitions\,in\,Abelian lattice\,gauge\,theories,\,{\it Nucl.\,Phys.}\,B\,{\bf 129},\,493--510 (1977).

%\bibitem{monjja}
%C.\,A.\,Trugenberger,\,M.\,C.\,Diamantini,\,N.\,Poccia,\,F.\,S.\,Nogueira, V.\,M.\,Vinokur,\,Magnetic\,monopoles and superinsulation\,in Josephson\,junction\,arrays.\,{\it Quantum\,Reports}\,{\bf 2},\,388-399\,(2020).

%\bibitem{moncon}
%M.\,C.\,Diamantini,\,C.\,A.\,Trugenberger,\,V.\,M.\,Vinokur,\,Quantum magnetic\,monopole\,condensate,\,{\it Comm. Physics}\,{\bf 4}, 25 (2021).


%\bibitem{pseudo}M.\,C.\,Diamantini, C.\,A.\,Trugenberger, V.\,M.\,Vinokur, Topological Nature of High Temperature Superconductivity, {\it Advanced Quantum Technologies} 2000135 (2021).

%\bibitem{parra} C.\,Parra, F.\,Niemstemski, A.\,W.\,Contryman, P.\,Giraldo-Gallo, T.\,H.\,Geballe,\,I.\,R.\,Fisher,\,and\,H.\,C.\,Manoharan,\,Signatures\,of two-dimensional superconductivity emerging within a three-dimensional\,host\,superconductor,\,{\it PNAS}\,{\bf 118},\,e2017810118\,(2021).




	
	
	
\bibitem{dirac} P. A. M. Dirac, Quantized singularities in the electromagnetic field, {\it Proc. Roy. Soc. (London)}\,A\,{\bf 133}, 60 (1931). 

\bibitem{olive} P. Goddard \& D. I. Olive, Magnetic monopoles in gauge field theories. {\it Rep. Prog. Phys.} {\bf 41}, (1978) 91. 




	
\bibitem{mandelstam}Mandelstam,\,S. Vortices and quark confinement in non-Abelian gauge theories. {\it Phys. Rep.} {\bf 23}, 245-249 (1976). 

\bibitem{nambu}Y. Nambu, Strings, monopoles and gauge fields. {\it Phys. Rev.}\,D\,{\bf 10}, 4262 (1974).

\bibitem{thooft} G. `t Hooft, On the phase transition towards permanent quark confinement. {\it Nucl. Phys.}\,B\,{\bf 138}, 1-25 (1978). 

\bibitem{greensite}J. Greensite, {\it An introduction to the confinement problem.} Springer-Verlag (Berlin) 2011. 



\bibitem{doniach} A. Kr\"amer and S. Doniach {\it Phys. Rev. Lett.} {\bf 81}, 3523 (1998). 


\bibitem{bat_vin_2007}
T.\,I.\,Baturina, A.\,Yu.\,Mironov, V.\,M.\,Vinokur, M.\,R.\,Baklanov, and C.\,Strunk, Localized Superconductivity in the Quantum-Critical Region of the Disorder-Driven Superconductor-Insulator Transition in TiN Thin Films. \textit{Phys. Rev. Lett.}\,\textbf{99}, 257003 (2007).

\bibitem{shahar} G. Sambandamurthy, L. W. Engels, A. Johansson, E. Peled \& D. Shahar, { Experimental evidence for a collective insulating state in two-dimensional superconductors} {\it Phys. Rev. Lett.} {\bf 94}, 017003 (2005). 


\bibitem{fistul}
M. V. Fistul, V. M. Vinokur \& T. I. Baturina, Collective Cooper-pair transport in the insulating state of Josephson-junction arrays. {\it Phys. Rev. Lett.} {\bf 100}, 086805 (2008).



\bibitem{ournodisorder} M.\,C.\,Diamantini, C.\,A.\,Trugenberger and V.\,M\,Vinokur, Superconductor-to-insulator transition in absence of disorder. {\it Phys. Rev.}\,B\,{\bf 103}, 174516 (2021). 



\bibitem{electrostatics}M. C. Diamantini, L. Gammaitoni, C. Strunk, S. V. Postolova, A. Yu. Mironov, C. A. Trugenberger and V. M. Vinokur, Direct probe of the interior of an electric pion in a Cooper pair superinsulator. {\it Nat. Comm. Phys.} {\bf 3}, 142 (2020). 


\
\bibitem{anyons} F. Wilczek, {\it Fractional statistics and anyon superconductivity}. World Scientific, Singapore (1990). 



 
 \bibitem{jackiw3} S. Deser, R. Jackiw \& S. Templeton, Topologically massive gauge theories. {\it Ann. Phys.} (N.Y.) {\bf 140}, 372-411 (1982). 



%\bibitem{das2}D.\,Das,  S.\,Doniach, Bose metal: gauge field fluctuations and scaling for field-tuned quantum phase transitions. {\it Phys. Rev.} B\,{\bf 64} 134511 (2001).

%\bibitem{phillips1}D.\,Dalidovich, P.\,Phillips, Phase Glass is a Bose Metal: A New Conducting State in Two Dimensions. 
%\textit{Phys. Rev. Lett}. \textbf{89}, 27001 (2002); 

%\bibitem{phillips2}P.\,Pillips, D.\,Dalidovich,  The Elusive Bose Metal. 
%{\it Science} {\bf 302} 243 -- 247 (2003).

%\bibitem{kapitulnik} A.\,Kapitulnik, S.\,A.\,Kivelson, and B.\,Spivak, Anomalous metals -- failed superconductors. \textit{Rev. Mod. Phys.}  \textbf{91}, 011002 (2019).



\bibitem{mironov}A. Yu. Mironov, D. M. Silevitch, T. Proslier, S. V. Postolova, M. V. Burdastyh, A. K. Gutakovskii, T. F.  Rosenbaum, V. M. Vinokur \& T. I. Baturina, Charge Berezinskii-Kosterlitz-Thouless transition in superconducting NbTiN films.
 {\it Scientific Reports} {\bf 8} 4082 (2018). 

\bibitem{sit1} K. B. Efetov, Phase transition in granulated superconductors. \textit{Sov. Phys. JETP} \textbf{51}, 1015 -- 1022 (1980); 

\bibitem{sit2} D. Haviland, Y. Liu, A. Goldman, Onset of superconductivity in the two-dimensional limit. \textit{Phys. Rev. Lett}. \textbf{62}, 2180 -- 2183 (1989); 

\bibitem{sit3} A.\,Hebard\,and\,M.\,A.\,Paalanen,\,Magnetic-field-tuned superconductor-insulator\,transition in two-dimensional films. \textit{Phys. Rev. Lett}. \textbf{65}, 927 -- 930 (1990);  


\bibitem{coleman} S. Coleman, {\it Aspects of symmetry}. Cambridge University Press, Cambridge (1985). 

\bibitem{poccia} C. A Trugenberger, M. C. Diamantini, N. Poccia, F. S. Nogueira \& V. M. Vinokur, Magnetic monopoles and superinsulation in Josephson junction arrays. {\it Quantum Reports} {\bf 2} 388-399 (2020). 

\bibitem{greensite}J. Greensite, {\it An introduction to the confinement problem.} Springer-Verlag (Berlin) 2011. 

\bibitem{nogueira}F. S. Nogueira, H. Kleinert H. Compact quantum electrodynamics in 2+1 dimensions and spinon confinement: a renormalization group analysis. {\it Phys. Rev.} \,B\,{\bf 77} 045107 (2008).

\bibitem{caselle}M. Caselle, M. Panero, D. Vadacchino, Width of the flux tube in compact U(1) gauge theory in three dimensions. {\it JHEP} {\bf 02}, 180 (2016). 

\bibitem{gross}D.\,Gross, Twenty Five Years of Asymptotic Freedom, {\it Nucl. Phys. B: Proceedings Supplements} {\bf 74}, 426 -- 446 (1998).

\bibitem{3Dgranular} C. Parra, F. Niemstemski, A. W. Contryman, P. Giraldo-Gallo, T. H. Geballe, I. R. Fisher and H. C. Manoharan, Signatures of two-dimensional superconductivity emerging within a three-dimensional host superconductor. {\it PNAS} {\bf 118} e2017810118 (2021). 


\bibitem{moncond} M. C. Diamantini, C. A. Trugenberger, V. M. Vinokur, Quantum magnetic monopole condensate, {\it Nature Comm. Phys. } {\bf 4} 25 (2021). 

\bibitem{axion}F. Wilczek, Two applications of axion electrodynamics. {\it Phys. Rev. Lett.} {\bf 58}, 1799 (1987). 

\bibitem{metlitski} M. A. Metlitski, C. L. Kane, \& M. P. A. Fisher, Bosonic topological insulator in three dimensions and the statistical Witten effect. {\it Phys. Rev.}\,B\,{\bf 88} 035131 (2013). 

\bibitem{senthil3D} 
A. Vishwanath, \& T. Senthil, Physics of Three-Dimensional Bosonic Topological Insulators:  Surface-Deconfined Criticality and Quantized Magnetoelectric Effect. {\it Phys. Rev. }\,X\,{\bf 3}, 011016 (2013).

\bibitem{witteneffect} E. Witten, Dyons of Charge $\theta /2\pi$. {\it Phys. Lett.} {\bf 86}, 283-287 (1979). 

\bibitem{highT1} M. C. Diamantini, C. A. Trugenberger and V. M. Vinokur,  Topological nature of high-temperature superconductivity, {\it Adv. Quantum Technologies} 2000135 (2021). 

\bibitem{revhighT} X. Zhou, W.-S. Lee, M. Imada, N. Trivedi, P. Phillips, H.-Y. Kee, P. T\"orm\"a \& M. Eremets, High-temperature superconductivity, {\it Nature Review Physics} {\bf 3} 462-465 (2021). 

\bibitem{strongin} M.\,Strongin, R.\,S.\,Thomson, O.\,F.\,Kammerer \& J.\,E.\,Crow, Destruction of superconductivity in disordered near-monolayer films. {\it Phys. Rev.}\,B\,{\bf 1} 1078-1091 (1970); 	
	
\bibitem{finkelstein} A.M. Finkel'shtein, Superconducting transition temperature in amorphous films. {\it JETP Lett. } {\bf 45} 46-49 (1987). 

\bibitem{trivedi}N. Trivedi, R. T. Scalettar \& M. Randeria, Superconductor-insulator transition in a disordered electronic system, {\it  Phys. Rev.}\,B\,{\bf 54} R3756 (1996). 

\bibitem{altshuler1} D. M. Basko, I. L. Aleiner \& B. L. Altshuler, Possible experimental manifestations of the many-body localization, {\it Phys. Rev. }\,B\,{\bf 76} 052203 (2007). 

\bibitem{altshuler2} B. L. Altshuler, V. E. Kravtsov, I. V. Lerner and I. L. Aleiner, Jumps in current-voltage characteristics in disordered films, {\it Phys. Rev. Lett.} {\bf 102} 176803 (2009).

\bibitem{jain}S. Pu, G. J. Sreejith \& J. K. Jain, Anderson localization inf the fractional quantum Hall effect, arXiv: 2109.00362.

\bibitem{kivelson} K.-S. Kim \& S. A. Kivelson, The quantum Hall effect in the absence of disorder, {\it Quantum Matter} {\bf 6} 22 (2021). 

\bibitem{sondhi1} R. M. Nandkishore and S. L. Sondhi, Many-body localization with long-range interactions, {\it Phys. Rev.} {X7} 041021 (2017).

\bibitem{sondhi2} A. A. Akhtar, R. M. Nandkishore and S. L. Sondhi, Symmetry breaking and localization in a random Schwinger model with commensuration, {\it Phys. Rev. }\,B\,{\bf 98} 115109 (2018). 


\fi


	
\end{thebibliography}
\end{document}